\documentclass[]{emulateapj}
\usepackage[breaklinks,colorlinks,citecolor=blue,linkcolor=magenta]{hyperref} 
\usepackage{amsgen,amsmath,amstext,amsbsy,amsopn,amssymb}
\usepackage{epstopdf}
\usepackage{graphics}
\usepackage[usenames]{color}
\usepackage{epsfig}
\usepackage{longtable}
\usepackage{bm} 
\usepackage{ulem}
\usepackage{url}
\usepackage{fontenc}
\usepackage{inputenc}
\usepackage{xcolor}
\usepackage{varwidth}
\usepackage{braket}

\pdfoutput = 1

\def\lsim{~\rlap{$<$}{\lower 1.0ex\hbox{$\sim$}}}
\def\bsim{~\rlap{$>$}{\lower 1.0ex\hbox{$\sim$}}}

\def\la{\langle}
\def\ra{\rangle}

\def\ln{{\rm ln}}

\def\bk{{\bf k}}
\def\bq{{\bf q}}

\def\be{\begin{equation}}
\def\ee{\end{equation}}
\def\bea{\begin{eqnarray}}
\def\eea{\end{eqnarray}}
\def\ba{\begin{align}}

\def\bi{\begin{itemize}}
\def\ei{\end{itemize}}

\def\vk{\mathrm{\bf k}}

\def\apjl{Astrophys.\ J.\ Lett.}

\def\aap{Astron.\ Astrophys.}

\def\apjs{Astrophys.\ J. Supp.}

\shorttitle{PNG from Line Intensity Mapping}
\shortauthors{Moradinezhad Dizgah $\&$ Keating}

\begin{document}

\title{Line intensity mapping with [CII] and CO(1-0) as probes of primordial non-Gaussianity}
\author{Azadeh Moradinezhad Dizgah$^{1}$$^{\ast}$ $\&$ Garrett K. Keating$^{2}$$^{\ast \ast}$}
\affil{\small $^{1}$Department of Physics, Harvard University, 17 Oxford St., Cambridge, MA 02138, USA}
\affil{\small $^{2}$Harvard-Smithsonian Center for Astrophysics, 60 Garden Street, Cambridge, MA 02138, USA}
\email{$^{\ast}$amoradinejad@physics.harvard.edu} 
\email{$^{\ast \ast}$garrett.keating@cfa.harvard.edu}

\begin{abstract}

Constraints on primordial non-Gaussianity (PNG) will shed light on the origin of primordial fluctuations and the physics of the early universe. The intensity mapping technique is a promising probe of structure formation on large scales; at high redshifts, it can provide complementary information to other cosmological probes. We explore the potential of future wide-field [CII] and CO intensity mapping surveys in constraining PNG of the local shape, which induces a distinct, scale-dependent correction to the line bias. We explore the parameter space of CO and [CII] survey designs that can achieve the nominal target of $\sigma(f_{\rm NL}^{\rm loc})  =1$, and further calculate what constraints such surveys can place on PNG of the equilateral, orthogonal, and quasi-single field shapes. We further test the dependence of these constraints on various model assumptions; namely the halo mass function, modeling of line bias, and correlation between mass and [CII]/CO luminosity.  We find that the ability of CO and [CII] intensity mapping surveys to constrain $f_{\rm NL}^{\rm loc}$ relies heavily on the spectral (redshift) coverage, requiring at least a coverage of an octave in frequency, to produce significant results, with the optimized surveys generally covering between $z\sim[2-8]$. As this redshift window partially overlaps with 21-cm EoR experiments like the Hydrogen Epoch of
Reionization Array (HERA), we additionally explore the prospects for a hypothetical set of surveys to place constraints on $\sigma(f_{\rm NL}^{\rm loc})$ utilizing multi-tracer analysis between the lines of HI, CO, and [CII].
\end{abstract}

\keywords{early universe ---  galaxies: high-redshift --- large-scale structure of universe}

\section{Introduction}\label{sec:intro}
 Understanding the origin of primordial fluctuations is one of the main open questions in cosmology. Current cosmological observations \citep[e.g.,][]{2015PhRvL.114j1301B,2016A&A...594A..20P} are compatible with the simplest models of inflation predicting adiabatic, super-horizon, scale-invariant, and nearly Gaussian primordial perturbations. High-precision constraints on the level of primordial non-Gaussianity (PNG) provides a unique window into the interactions of quantum fields during inflation, enabling one to distinguish between multitude of inflationary models. Currently, the best constraints on PNG are those from the Cosmic Microwave Background (CMB) data from Planck satellite \citep{2016A&A...594A..17P}, obtained by measuring the CMB temperature and polarization bispectra. Going beyond the precision achieved by CMB will likely be possible with large scale structure (LSS) surveys through measurements of scale-dependent bias \citep{Dalal:2007cu,Matarrese:2008nc,Afshordi:2008ru} and higher-order correlation functions \citep{Scoccimarro:2003wn,Sefusatti:2007ih}. 

Current and future galaxy surveys such as BOSS \citep{2017MNRAS.470.2617A}, DES \citep{2017arXiv170801530D}, DESI \citep{2016arXiv161100036D}, EUCLID \citep{2018LRR....21....2A} and LSST \citep{2009arXiv0912.0201L} probe the underlying matter distribution by resolving individual biased tracers, such as galaxies and quasars, to make a three-dimensional map of the Universe. In contrast, line intensity mapping probes the large-scale matter distribution by measuring the cumulative light from an ensemble of sources, including faint galaxies not resolved in galaxy surveys, while preserving accurate redshift information. The aggregate emission from galaxies is observed as fluctuations in the mean line intensity (or equivalently, brightness temperature). Therefore, it has a great potential to probe the Universe at redshifts and scales beyond what is accessible with any galaxy surveys.  

In addition to the 21-cm hyperfine emission of neutral hydrogen \citep[e.g.,][]{Pritchard:2011xb, Barkana:2016nyr}, rotational lines of carbon monoxide (CO) \citep{visbal:2010rz,Lidz:2011dx,Carilli:2011vw,Pullen:2012su,Breysse:2014uia,Mashian:2015his, Li:2015gqa,Padmanabhan:2017ate}, the fine-structure line from ionized carbon ([CII]) \citep{Gong:2011mf,Silva:2014ira,Yue:2015sua,Lidz:2016lub,Pullen:2017ogs} and the Ly-$\alpha$ \citep{Silva:2012mtb,Pullen:2013dir} emission line are some of the most widely studied intensity mapping candidates. Both the $J_{1\rightarrow 0}$ transition of CO ($\nu_{\rm rest}=115.271$ GHz), herein referred to as CO(1-0), and the [CII] 158-$\mu$m fine structure line ($\nu_{\rm rest } = 1900.539$ GHz) are interesting candidates for line intensity mapping studies, particularly as  emission from both lines at high redshifts is generally accessible from the ground. CO is the second most abundant molecule in interstellar medium after molecular hydrogen, and [CII] is typically the brightest spectral line of star-forming galaxies, in the far-infrared (FIR) \citep{Carilli:2013qm}. On small scales, CO and [CII] typically trace star-forming galaxies -- more specifically, the cool gas within these galaxies that will provide the fuel for star formation. However, on large scales these lines map the distribution of star-forming galaxies over a wide range of redshifts and can be used as tracers of large scale structure to constrain cosmology. CO and [CII] have been detected in individual galaxies and quasars at high redshift \citep{Carilli:2013qm,2015MNRAS.452...54M,Walter:2016co,2018ApJ...864...49P}. 

Recent observation work has resulted in the first tentative power spectrum detections of both CO and [CII].  \cite{2016ApJ...830...34K} reported a measurement of the CO power spectrum intensity mapping at redshift $2.3\leq z\leq3.3$, while [CII] was tentatively detected via cross-correlation between Planck and SDSS (II and III) spectra by \cite{Pullen:2017ogs}. These early studies are complemented by several upcoming ground-based dedicated intensity mapping surveys such as COMAP \citep{Li:2015gqa} for CO and TIME \citep{doi:10.1117/12.2057207}, CCAT-prime, CONCERTO \citep{Serra:2016jzs} for [CII]. Moreover, PIXIE \citep{2011JCAP...07..025K}, the proposed NASA Medium-class Explorer mission, has the capability of high-precision measurement of [CII] over full-sky. There are still many uncertainties in theoretical modeling of the power spectrum of emission lines. However given its potential and multitude of upcoming surveys dedicated to intensity mapping with CO and [CII], it is worth exploring the science potential.

In the presence of PNG, one needs to account for the higher-order correlation functions in addition to the 2-point function (or power spectrum in Fourier space) to describe the statistics of primordial fluctuations. At lowest order, the three-point function -- or its Fourier equivalent, the bispectrum -- captures additional information about PNG not captured by the power spectrum. Commonly, the primordial bispectrum is parametrized in terms of an overall amplitude $f_{\rm NL}$ and a shape function. Different models of inflation give rise to different shapes of bispectrum. Therefore, constraints on the amplitude of a given shape sets constraints on the physical mechanism at play during inflation. A non-zero primordial bispectrum leaves a signature on the power spectrum of biased tracers, by inducing a correction to the bias, which can be used to constrain $f_{\rm NL}$. For the local shape in particular, the correction has a strong $1/k^2$ scale-dependence \citep{Dalal:2007cu,Matarrese:2008nc}. Since this signal is not commonly sourced by other astrophysical sources, it is considered a clean probe of local shape PNG and has been already used to obtain constraint on $f_{\rm NL}^{\rm loc}$ from redshift surveys \citep{Slosar:2008hx,2013MNRAS.428.1116R,Giannantonio:2013uqa,Giannantonio:2013kqa,2015JCAP...05..040H,Leistedt:2014zqa}.  

As emission lines are biased tracers of underlying matter distribution, the presence of PNG induces a correction to the line bias. In this paper, we explore the potential of future [CII] and CO intensity mapping surveys in constraining PNG via the measurement of the clustering power spectrum. We focus on PNG of the local shape \citep{Gangui:1993tt,Wang:1999vf, Verde:1999ij, Komatsu:2001rj}, since its signature on the power spectrum is the most distinct, but we also explore other shapes; namely equilateral \citep{Creminelli:2005hu,Babich:2004gb}, orthogonal \citep{Senatore:2009gt}, and quasi-single field \citep{Chen:2009zp} models. We first study what are the requirements for a survey that can achieve the target sensitivity of $\sigma(f_{\rm NL}^{\rm loc}) = 1$, as a function of several parameters, including integration time, survey area, and redshift coverage. This is a theoretically motivated target, since all single-field models of inflation in the attractor regime are expected to produce $f_{\rm NL}^{\rm loc} \ll 1$ \citep{Maldacena:2002vr,Creminelli:2004yq}. As the local shape is a sensitive probe of multifield inflation, achieving such a precision in the measurement of local non-Gaussianity would enable us to distinguish between the two scenarios. Furthermore, on very large scales the general relativistic projection effects \citep{Yoo:2010ni,Bonvin:2011bg, 2011PhRvD..84d3516C, Jeong:2011as,Bruni:2011ta} also give rise to an effective scale-dependent correction to the linear bias similar to that due to local PNG with $f_{\rm NL}^{\rm loc} = \mathcal{O} (1)$, providing an ``observational floor'' for the measurement of large-scale scale-dependent bias. Having characterized ``optimal instruments'' for achieving the target sensitivity on measurement of $f_{\rm NL}^{\rm loc}$, we then study how the constraints from such a survey depend on the astrophysical model assumptions. We further discuss the impact of the foregrounds and discuss the scale of the required survey and instrument design in the context of planned and upcoming intensity mapping surveys targeting CO and [CII] lines.  

The rest of the paper is organized as follows: in Section \ref{sec:PS_CO} we review the theoretical model of the power spectrum of the line intensity. Next, in Section \ref{sec:survey_design}, we describe the instrument and survey attributes that we probe in order to determine the optimal specifications for constraining PNG of the local shape. After detailing our forecasting methodology in Section \ref{sec:forecast}, we present our results in Section \ref{sec:results}. Based on these results, we discuss the impact of the foregrounds and the experimental landscape in Section \ref{sec:discussion}, and draw our conclusions in Section \ref{sec:conclusion}.

\section{Power spectrum of line intensity}\label{sec:PS_CO}\
Here we discuss our model of the power spectrum of line intensity in the presence of PNG. These models have primarily built on existing models in the literature, accounting for redshift-space distortions (RSD) and the Alcock-Paczynski (AP) effect. In order to test the dependence of model properties on the constraining power for a given survey, we generate both fiducial and variant models for CO(1-0) and [CII] line emission, for which constraints are separately calculated for in Section \ref{ssec:results_astro_model}.

\subsection{Power Spectrum for Gaussian Initial Conditions}\label{ssec:PS_G} 
We use a simple model to describe the power spectrum of line emission from galaxies over a wide redshift range,  which relates the mean intensity of the emission line to the abundance of halos that host CO- or [CII]-luminous galaxies \citep{visbal:2010rz,Gong:2011ts,Lidz:2011dx,Silva:2014ira} (see \cite{2016MNRAS.461...93P,2018MNRAS.473..271V,2019MNRAS.482.4906P,2018A&A...609A.130L} and references therein for more detailed modeling of CO and [CII], based on semi-analytical models in combination with hydrodynamical simulations). 

The mean brightness temperature (typically in units of $\mu K$) at redshift $z$ is given{} by 
\begin{equation}
\langle T_{\rm line}\rangle (z)  = \frac{c^2 f_{\rm duty}}{2k_B \nu_{\rm obs}^2} \int_{M_{\rm min}}^{M_{\rm max}} dM \frac{dn}{dM} \frac{L(M,z)}{4 \pi \mathcal{D}_{L}^{2}} \left ( \frac{dl}{d\theta} \right )^{2} \frac{dl}{d\nu},
\end{equation}
where $f_{\rm duty}$ is the duty cycle of line emitters defined as a fractional time the halo hosts a CO- or [CII]-luminous galaxy, or alternatively, the fraction of halos with mass larger than $M_{\rm min}$ that actively emit a given line at a given redshift. $dn/dM$ is the halo mass function (i.e. number of halos per unit mass at a given redshift), $L(M,z)$ is the specific luminosity of the galaxy located in a halo of mass $M$ at redshift $z$ luminous in a given line, and  ${\mathcal D}_L$ is the luminosity distance. The terms $dl/d\theta$ and $dl/d\nu$ reflect the conversion from units of comoving lengths, $l$, to those of the observed specific intensity: frequency, $\nu$, and angular size, $\theta$. The term $dl/d\theta$ is equivalent to comoving angular diameter distance, whereas 

\begin{equation}
\frac{dl}{d\nu} = \frac{c(1+z)}{\nu_{\rm obs}H(z)}, 
\end{equation}
where $H(z)$ is the Hubble parameter at a given redshift. Clustering of matter induces fluctuations in the brightness temperature of the line. 

On large scales the bias, $b_{\rm line}(z)$, between the line intensity and matter density fluctuations is assumed to be linear, and the clustering component of the line intensity power spectrum (typically in units of $\mu$K$^2$ Mpc$^{-3}$) can be expressed as
 \be\label{eq:ps_r}
 P_{\rm clust}(k,z) = \left[ \la T_{\rm line}\ra (z)\right]^2 b_{\rm line}^2(z) P_0(k,z),
 \ee
where $k$ is the comoving wavenumber, $P_0(k,z)$ is the linear dark matter power spectrum, and $b_{\rm line}(z)$ is the luminosity-weighted bias of the line intensity. Since CO or [CII] are emitted from within halos, their bias can be related to the bias of halos of mass $M$ at redshift $z$, $b_h(M,z)$ as 
\begin{equation}\label{eq:line_bias}
b_{\rm line}(z) = \frac{\int_{M_{\rm min}}^{M_{\rm max}} dM  \ \frac{dn}{dM} \  b_h(M,z) L(M,z)  
}{\int_{M_{\rm min}}^{M_{\rm max}} dM \ \frac{dn}{dM} \ L(M,z)},
\end{equation}
where $M_{\rm min}$ and $M_{\rm max}$ are the minimum and maximum mass of the CO- and [CII]-luminous halos.

An additional contribution to the observed power spectrum arises due to the fact that galaxies are discrete tracers of the underlying dark matter distribution, inducing shot noise in our measurement. Assuming galaxies are sampled from a Poisson process, this adds a  contribution to the line power spectrum which is given by the inverse number density of galaxies luminous in a given line, i.e., in every mass bin $\Delta M$, the added power amounts to  $\left(f_{\rm duty}\Delta  M dn/dM\right)^{-1}$. The shot power can be expressed as
\begin{equation}\label{eq:ps_shot}
P_{\rm shot}(z) = \frac{c^4 f_{\rm duty}}{4k_B^2 \nu_{\rm obs}^4}  \int_{M_{\rm min}}^{M_{\rm max}} dM \frac{dn}{dM} {\left[\frac{L(M,z)}{4 \pi \mathcal D_L^2} 
\left ( \frac{dl}{d\theta} \right )^{2} \frac{dl}{d\nu} \right ]}^2.
\end{equation} 
Note the shot-noise has linear dependence on $f_{\rm duty}$ while the clustering power spectrum has quadratic dependence.

Finally, the instrumental noise, discussed further in Section \ref{sec:survey_design}, plays an important role and affects the precision with which PNG can be extracted from the intensity mapping surveys. We defer the discussion of this component to Section \ref{ssec:inst_noise}.
 
\subsection{Astrophysical Model Variations} \label{ssec:model_var}
\noindent {\bf Intrinsic CO luminosity:} There have been two approaches in the literature in modeling the CO luminosity, the first uses empirical data to model the relation between CO and far-infrared (FIR) luminosities, the correlation of FIR luminosity and star-formation-rate (SFR) and the relation between the SFR and the halo mass \citep{Righi:2008br,visbal:2010rz,Lidz:2011dx,Pullen:2012su,Li:2015gqa}. The second approach considers the SFR required to reionize the neutral IGM at high redshift and subsequently converts the SFR to CO luminosity \citep{Gong:2011ts,Carilli:2011vw}. 

In our analysis, we follow the first approach and consider two models. As our base model, we use the model of \cite{Li:2015gqa}. This model uses the results of \cite{Behroozi:2012sp, Behroozi:2012iw} for ${\overline {\rm SFR}}(M,z)$,  the average SFR as a function of halo mass $M$ and redshift $z$. Note that the star formation rate of \cite{Behroozi:2012sp} only goes to $z\simeq8$. For higher redshifts $(z>8$), we extrapolate their result at lower redshift according to 
\begin{align}\label{eq:SFR_extrap}
\log \overline {\rm SFR}(M,z) = {\rm min}&\left[\overline {\rm SFR}(M,z=8) + 0.2943(z-8) \right. \nonumber \\
					&,\left. 3.3847-(0.2413z)\right]
\end{align}
The SFR (in units of $M_\odot {\rm yr}^{-1}$) is related to the total infrared luminosity $L_{\rm IR}$ (in units of $L_\odot$) via Kennicut relation  \citep{Kennicutt:1998zb} of the form
\begin{equation}
{\overline {\rm SFR}}(M,z) = \delta_{\rm MF} \times 10^{-10} L_{\rm IR}.
\end{equation}
The normalization depends on the assumptions of initial mass function, the duration of star-formation, etc. As in \cite{Behroozi:2012sp, Li:2015gqa}, we take $\delta_{\rm MF} =1$. The far-infrared luminosity $L_{\rm IR}$ (in units of $L_\odot$) is then related to the CO line luminosity $L'_{\rm CO}$ (in units of ${\rm K} \ {\rm km} s^{-1} {\rm pc}^2$), through a power-law fit of the form
\begin{equation}\label{eq:CO_lum}
\log L_{\rm IR} = \alpha \log  L'_{\rm CO} + \beta.
\end{equation}
As in the fiducial model of \cite{Li:2015gqa}, we use the fit from \cite{Carilli:2013qm} which found $\alpha = 1.37$ and $\beta = -1.74$ using a census of high redshift galaxies.  For CO(1-0), the line luminosity can then be expressed in units of solar luminosity ($L_{\rm CO(1-0)}$) via the following expression: 
\be
L_{\rm CO(1-0)} = 4.9 \times 10^{-5}  L'_{\rm CO}.
\ee

\begin{figure}
\centering
\includegraphics[width=0.45 \textwidth]{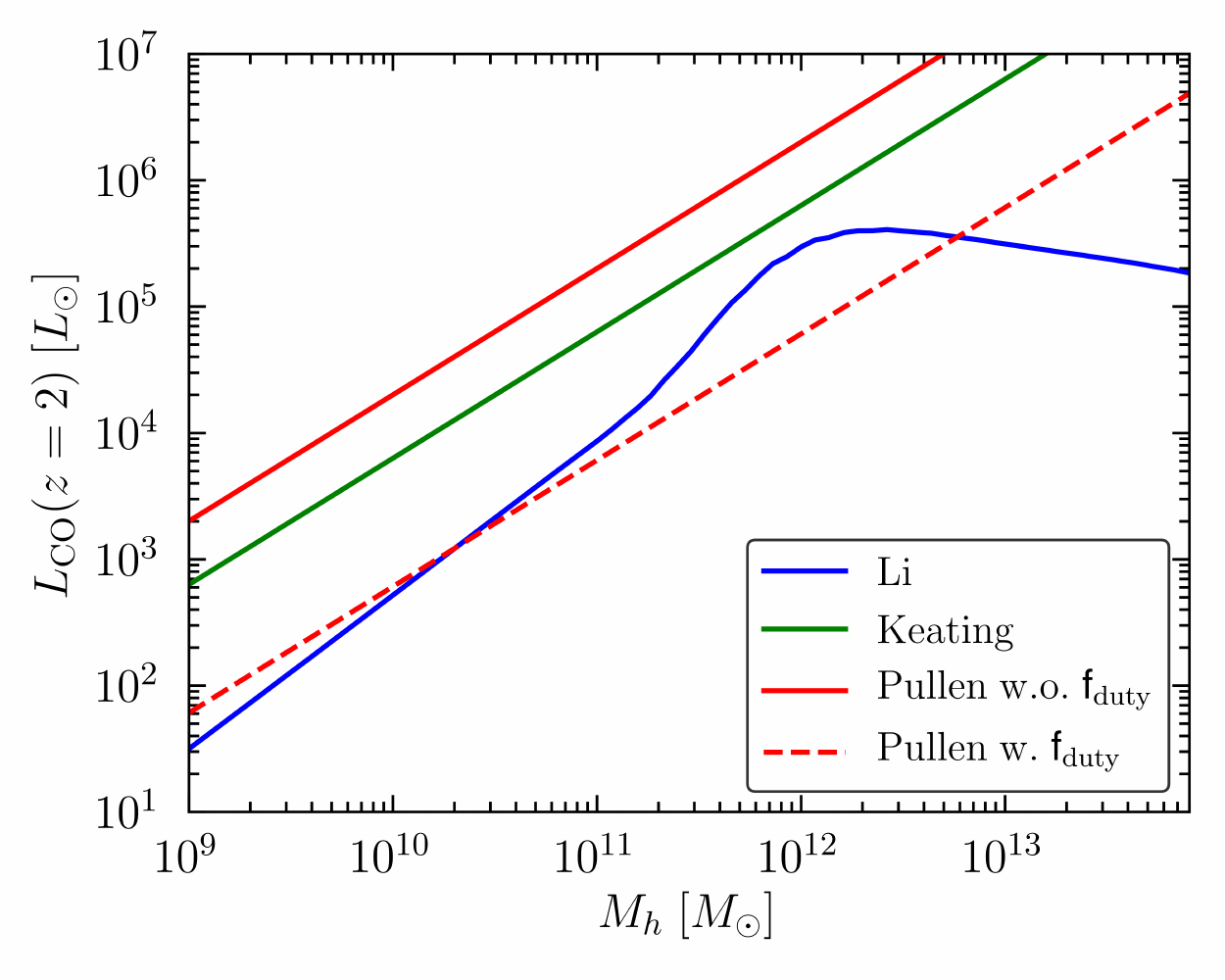}
\caption{The CO luminosity as a function of halo mass at $z=2$ for the models we consider in this work. The \citeauthor{Li:2015gqa} model is plotted without the log-scatter parameters $\sigma_{\rm CO}$ and $\sigma_{\rm SFR}$. The red solid line corresponds to model A of \citeauthor{Pullen:2012su} without multiplying by $f_{\rm duty}$ while the dashed red line shows this model accounting for the $f_{\rm duty}$.}
\label{fig:L_CO}
\end{figure}

The second set of models we consider assume a linear, redshift-independent relation between mass and specific luminosity $L_{{\rm CO}(1-0)} = A_{\rm CO} M $. These models also utilize the Kennicut relation to connect the FIR luminosity and SFR but assume a simple power-law relation between SFR and halo mass. For this linear model, we consider two values for the amplitude (in units of  $L_\odot M_\odot^{-1}$): $A_{\rm CO} = \{2 \times 10^{-6}, 6.3 \times 10^{-7}  \}$ corresponding to model A of \cite{Pullen:2012su}, and the value constraint by the measurement of the CO power spectrum by \cite{2016ApJ...830...34K}. In Figure \ref{fig:L_CO} we show the CO luminosity as a function of mass for $z=2$ for our base model and the two linear models. For the model of \citeauthor{Pullen:2012su}, we show the luminosity with and without the $f_{\rm duty}$ factor. 

\begin{figure}[htbp!]
\includegraphics[width=0.45 \textwidth]{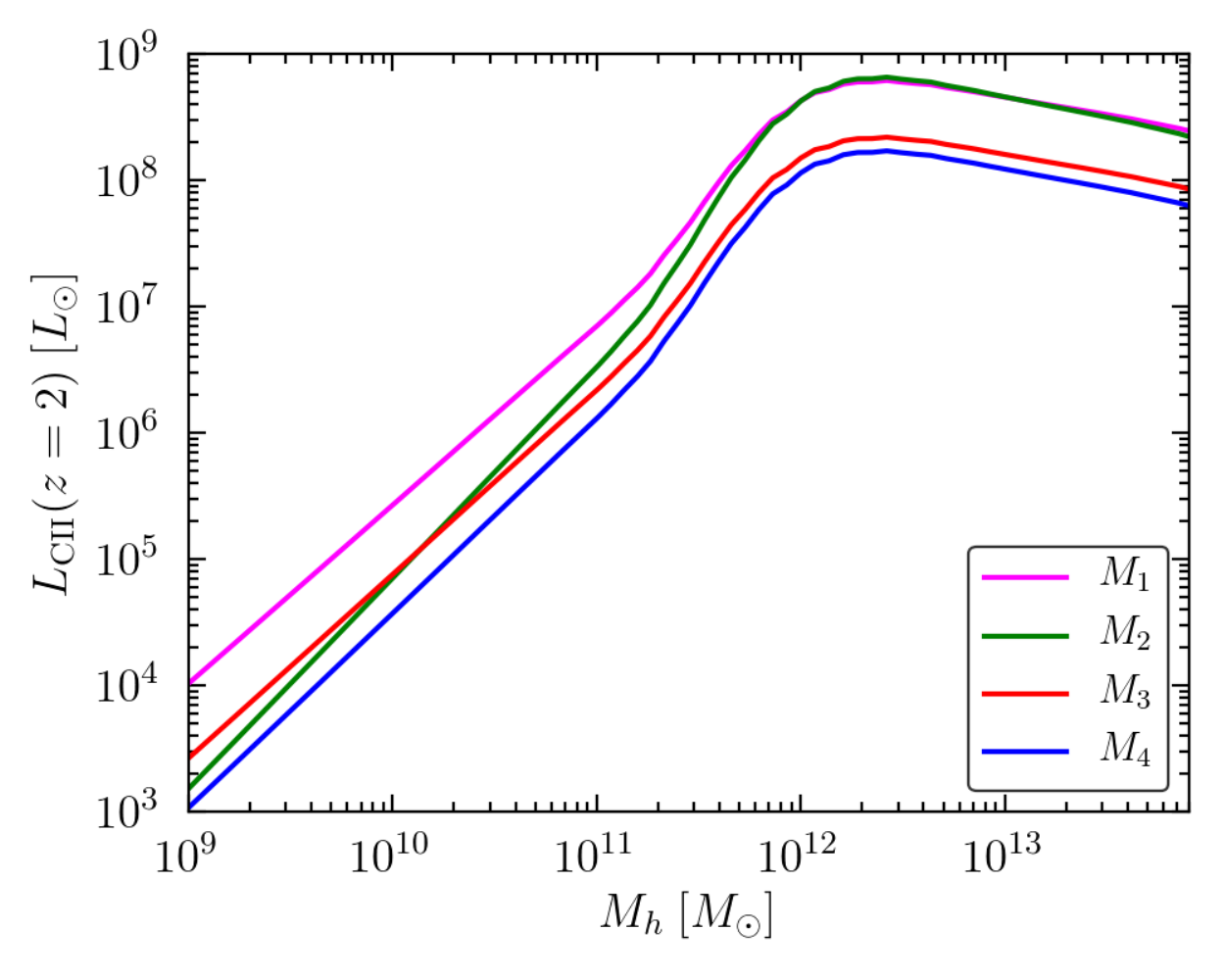}
\caption{The [CII] luminosity as a function of halo mass at $z=2$ for 4 models of  \cite{Silva:2014ira} that we consider in this work, labeled as $M_1,M_2,M_3,M_4$, as stated in the text.}
\label{fig:L_CII}
\end{figure}

\vspace{.1in}
\noindent {\bf Intrinsic [CII] luminosity:} 
[CII] emission is sourced from the colder diffuse media of galaxies (i.e., the cold and warm neutral media, as well as the photon-dominated regions around molecular clouds), and is well-correlated with the FIR emission of galaxies \citep{Stacey:1991cii}. In modeling the luminosity of [CII]-luminous galaxies, we use the results of \cite{Silva:2014ira} and consider the 4 models they presented in that work, labeled as $M_1,M_2,M_3,M_4$, which relate the [CII] luminosity to the star formation rate via a power-law scaling relationship, where
\be \label{eq:[CII]_lum}
\log L_{\rm CII} = a_{\rm LCII} \times \log \overline {\rm SFR}(M,z) +b_{\rm LCII},
\ee 
with the values of $a_{\rm LCII} $ and $b_{\rm LCII}$ given in Table \ref{tab:CII_models}. Similar to our model for ${\rm CO}_{(1-0)}$, we adopt the $\overline{\rm SFR}$ from \cite{Behroozi:2012sp}, utilizing the same extrapolation for $z>8$.  In Figure \ref{fig:L_CII} we show the [CII] luminosity as a function of mass for $z=2$ for the four models of \cite{Silva:2014ira}. 

\begin{table}[htbp!]
\caption{Parameters of the models of [CII] luminosity given in \cite{Silva:2014ira} that we consider in our forecast.}
\par\smallskip
\centering
\begin{tabular}{c c c  }
\hline \hline
${\rm Model}$ 	& $ a_{\rm LCII}$ & $b_{\rm LCII}$  \\  \hline 
$M_1$ & 0.8475 & 7.2203   \\
$M_2$ & 1.0000 & 6.9647  \\
$M_3$ & 0.8727 &  6.7250   \\
$M_4$ & 0.9231 & 6.5234 \\
\hline
\end{tabular}
\label{tab:CII_models} 
\end{table} 

\begin{figure}\centering
\includegraphics[width=0.45 \textwidth]{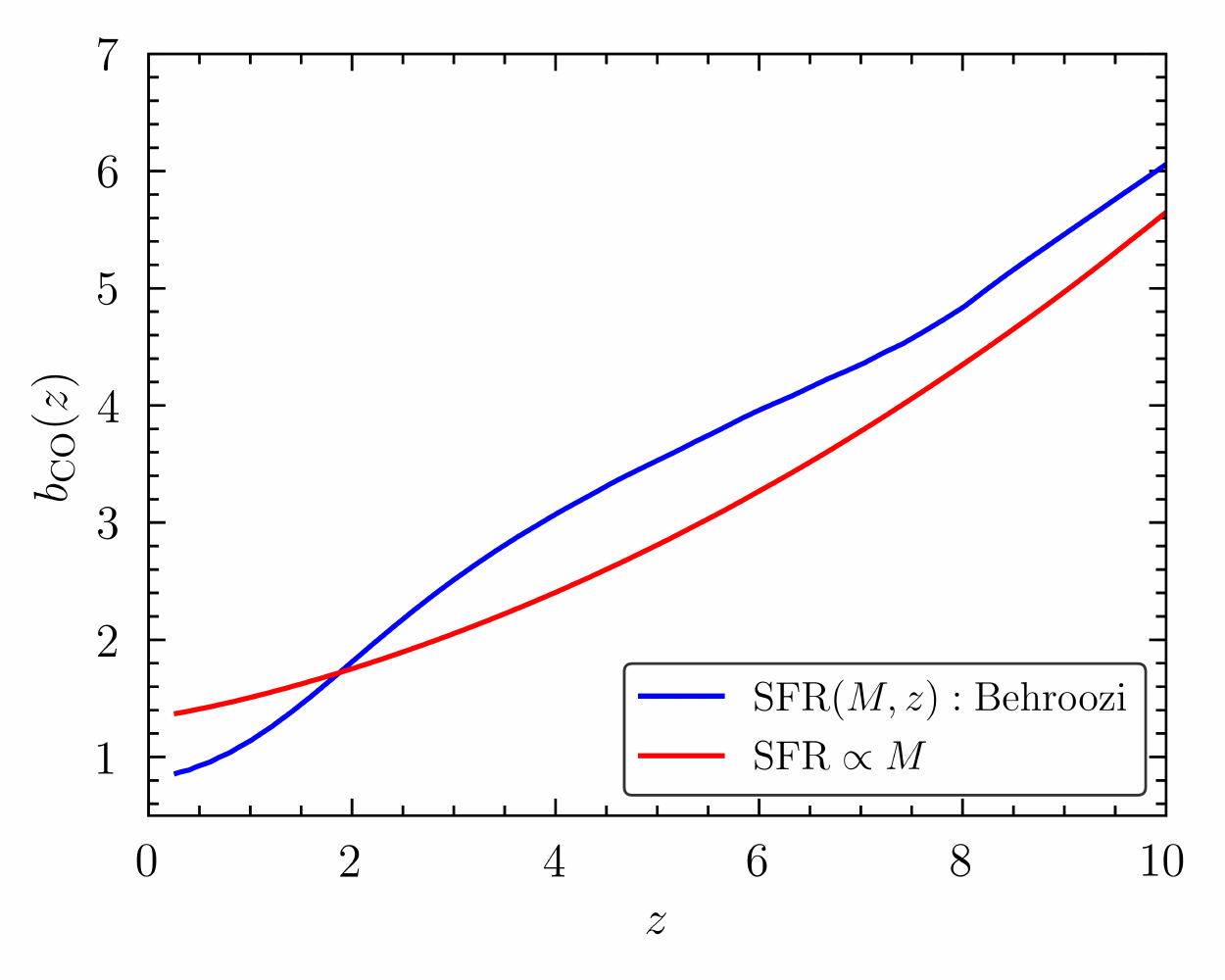}
\includegraphics[width=0.45 \textwidth]{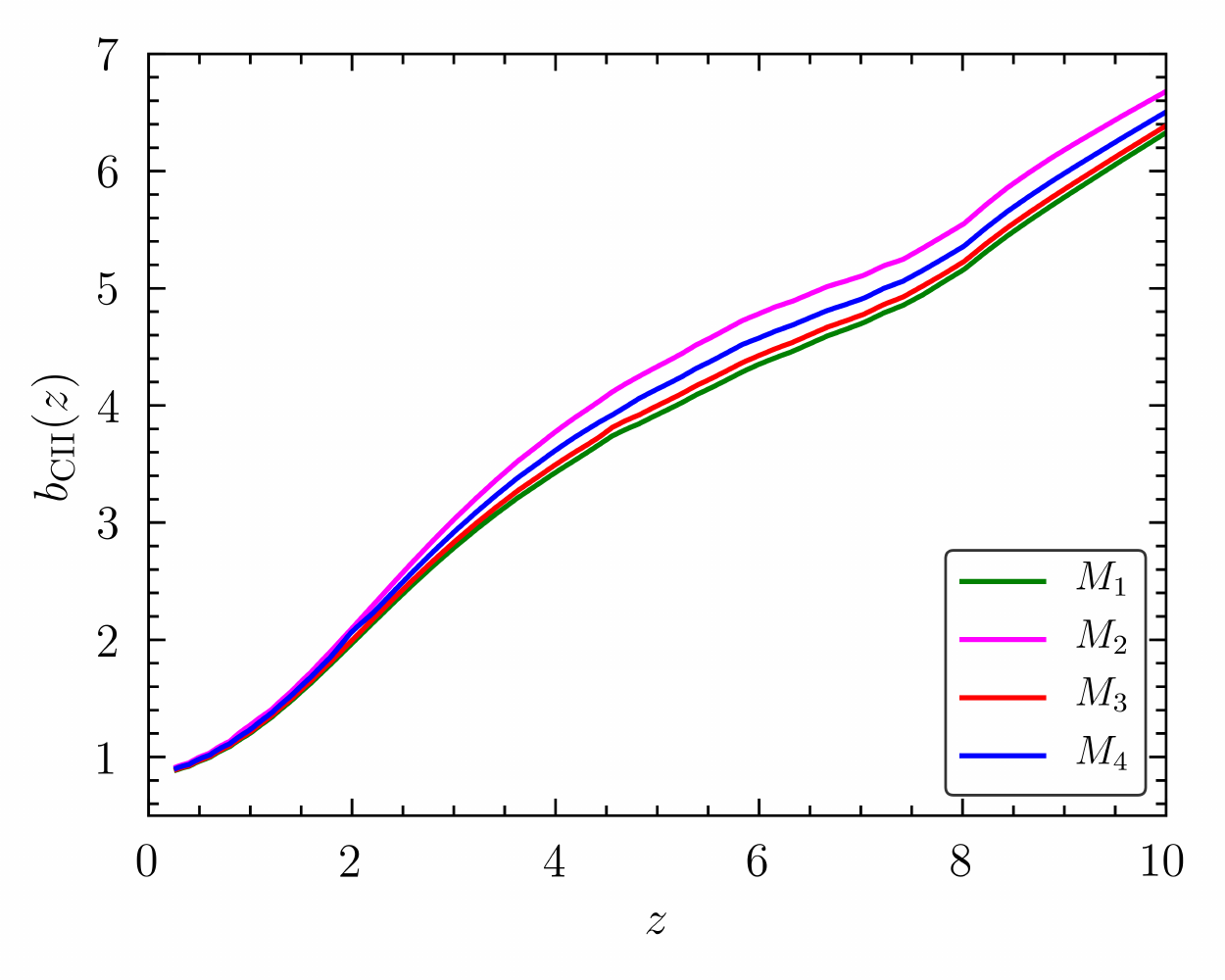}
 \caption{Biases of CO (top) and [CII] (bottom) as a function of redshift. Lines correspond to different models of specific luminosity as stated in the text.}
\label{fig:line_bias}
\end{figure}

In Figure \ref{fig:line_bias}, we show the CO and [CII] biases, calculated from Equation \eqref{eq:line_bias} for models of the specific luminosity as outlined in the next subsection. In the top panel, the red line corresponds to the model where the star formation rate is given by \cite{Behroozi:2012sp, Behroozi:2012iw}, while the blue line corresponds to the case where a linear relation between the SFR and halo mass is assumed. The bottom panel shows the bias for four models of [CII] luminosity given in \citep{Silva:2014ira}. We note that the small deflection at $z\approx 8$ is an artifact of the extrapolation used to extend the \cite{Behroozi:2012sp, Behroozi:2012iw} fit for the star-formation rate at a given halo mass to higher redshifts (as given in Equation \eqref{eq:SFR_extrap}), and is not physical. As this effect only exists over a limited redshift window, where mean brightness temperature is rapidly falling, and thus, not likely to contribute much to the constraining power of a given survey, we do not expect this artifact to significantly impact the analyses presented here.  

\vspace{.1in}
\noindent {\bf Duty cycle,} ${\mathbf f_{\rm duty}}$ {\bf versus log-scatter} ${\mathbf \sigma_{\rm line}}$ {\bf :} There is large uncertainty in modeling $f_{\rm duty}$, since it is poorly constrained observationally. Some models assume the star formation timescale, $t_s$, sets the duty cycle for molecular gas emission, such that $f_{\rm duty} = t_s/t_{\rm age}(z)$, where $t_{\rm age}(z)$ is the age of the universe \citep{visbal:2010rz,Lidz:2011dx,Pullen:2012su,Breysse:2014uia}. As the star-formation timescale of the order $t_s = 10^8 {\rm yr}$, at the redshift range of  $1<z<5$ for instance, the duty cycle is $0.017< f_{\rm duty}<0.085$. This choice of $f_{\rm duty}$ is made based on the assumption that at high redshifts, $z \gtrsim 6$,  galaxies undergoing extreme starburst-like events are the dominant source of CO emission \citep{Righi:2008br,Lidz:2011dx}. At lower redshifts of $z \sim (1-4)$, however, several observations \citep{Lee:2008yj,2007ApJ...660L..43N,2013ApJ...768...74T} indicated a near unity value of $f_{\rm duty}$, significantly larger than the values predicted by this model. For CO, \citet{Li:2015gqa} consider an alternative model to describe the observed variation. They adopt a pair of log-normal scatter parameters $\sigma_{\rm SFR}$ and $\sigma_{L_{\rm CO}}$ (which in principle can be combined into a single parameter) in the relation between star formation rate and specific luminosity with halo mass, while assuming $f_{\rm duty}\sim1$. \citet{2016ApJ...830...34K} utilize a variation of this model with a single, aggregate scatter parameter $\sigma_{\rm line}$.  Under this model, the $f_{\rm duty}$ terms in the mean brightness and shot power terms is replaced by
\be\label{eq:scatter}
p_{n,\sigma} = \int_{-\infty}^\infty dx \frac{10^{n x}}{\sqrt{2\pi}\sigma_{\rm line}} e^{-x^2/2\sigma_{\rm line}^2},
\ee
with $n=1$ for $\langle T_{\rm line} \rangle$, and $n=2$ for $P_{\rm shot}$. Note that $p_{n,\sigma}$ is a monotonically growing function of $\sigma_{\rm line}$, and is always equal or greater than unity, while $f_{\rm duty}\leq1$. In our forecast for CO with \citeauthor{Li:2015gqa} and \citeauthor{2016ApJ...830...34K} luminosities, we consider a single scatter parameter as in \cite{2016ApJ...830...34K} and set the value of  $\sigma_{\rm CO} = 0.37$, corresponding to fiducial model of \cite{Li:2015gqa}. For \citeauthor{Pullen:2012su}, we use the $f_{\rm duty}$ parametrization and we consider $f_{\rm duty} =   t_s/t_{\rm age}(z)$. For [CII], we will also only consider a single scatter parameter and set it to $\sigma_{[\rm CII]} = 0.37$.

\vspace{.1in}
\noindent {\bf Mass function and the line bias:} In our base model, we use the Sheth-Tormen (ST) \citep{Sheth:1999mn} mass function (assuming \cite{Sheth:1999su} halo bias). To study the dependence of our forecast on the choice of halo mass function we also consider the Press-Schecter (PS) \citep{Press:1973iz,Bond:1990iw} and Tinker (TR) \citep{Tinker:2010my} mass functions. In modeling the line bias, we consider two models, one where the values of biases at a given redshift are fixed to be that given by Equation \eqref{eq:line_bias}. In the second case, assuming the biases in each redshift bin to be independent of one another, we vary one free parameter for the value of bias in each redshift bin and marginalize over them to obtain constraints on PNG. In this case, we assume the fiducial values of the bias to be given by Equation \eqref{eq:line_bias}. 

\begin{figure}
\includegraphics[width=0.45 \textwidth]{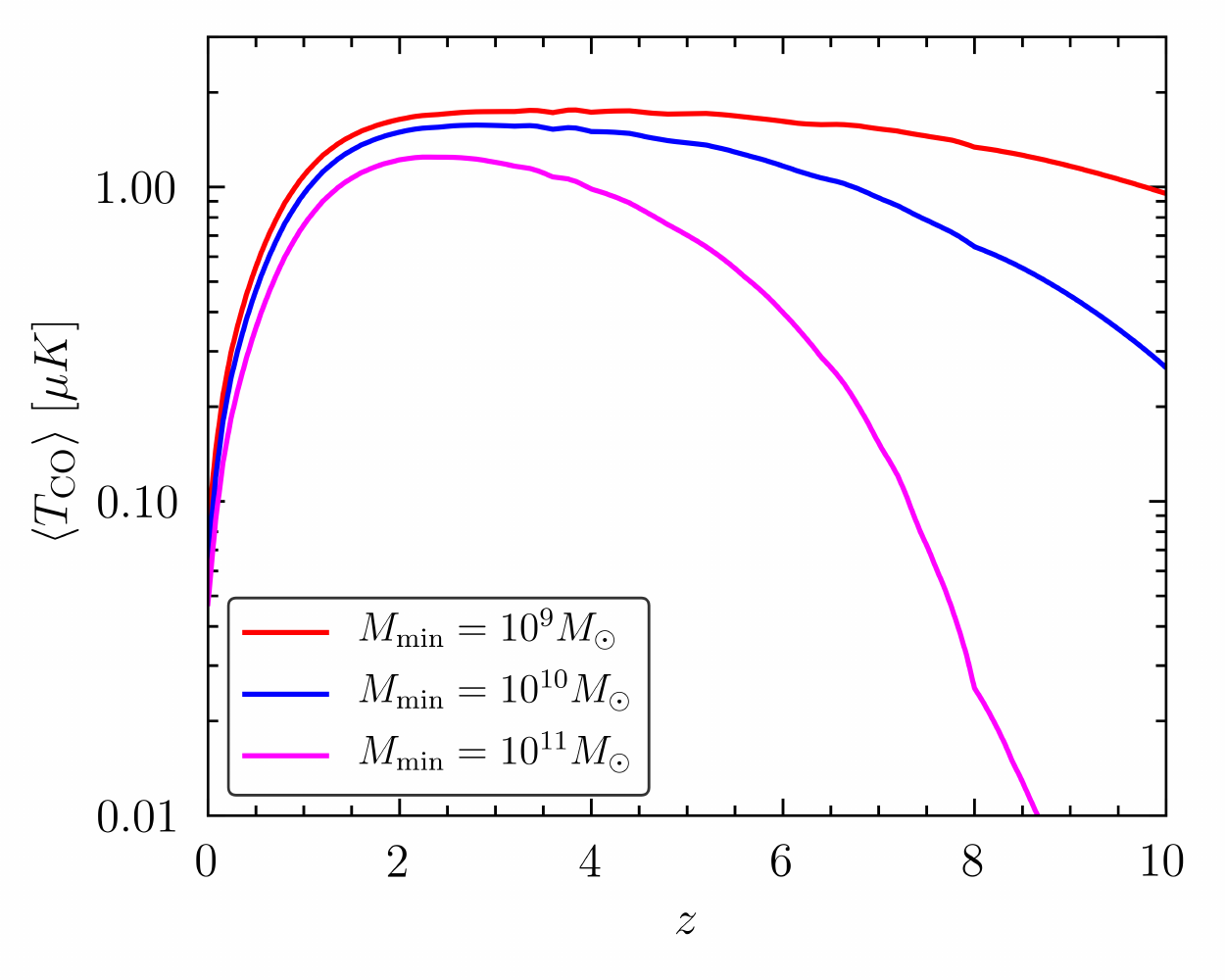}
\includegraphics[width=0.45 \textwidth]{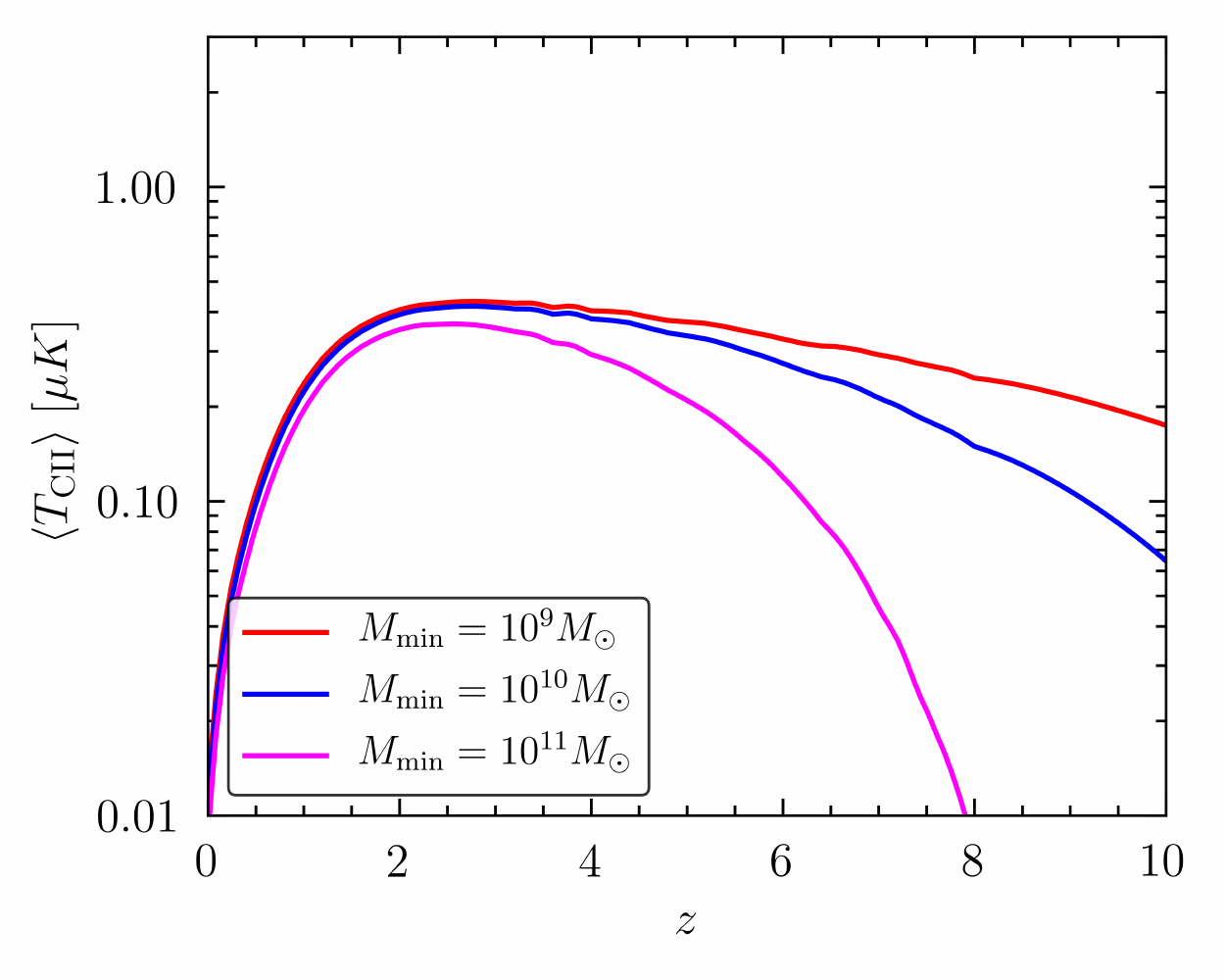}
\caption{Mean brightness temperature as a function of redshift, for three values of $M_{\rm min}$, for CO  (top) and CII (bottom). The CO luminosity is assumed to be given by \citet{Li:2015gqa}, while CII luminosity is taken to be M1 model of \citet{Silva:2014ira}. We accounted for the scatter in the mass-luminosity relation as in Equation \eqref{eq:scatter} and used the Sheth-Tormen halo mass function.}
\label{fig:Tbar_line}
\end{figure}

\vspace{.1in}
\noindent {\bf Minimum mass of [CII]- and CO-luminous halos:} In this work, $M_{\rm min}$ can be interpreted as the point at which the observed correlations between line emission and star-formation rate (Equations \ref{eq:CO_lum} and \ref{eq:[CII]_lum}) break down, such that galaxies with host halos below this mass do not significantly contribute to the global line emission. CO and [CII] emission are both sensitive to the gas metallicity of a given galaxy; CO particularly so, due to its reliance on dust for shielding from photo-dissociation \citep{Carilli:2013qm,Sargent:2014co,Olsen:2017cii}. This may be especially important for low-mass objects at high redshift, whose star formation histories and stellar masses are too limited to have molecular gas with enough chemical enrichment to show significant CO or [CII] line emission. We therefore use $M_{\rm min}$ as a proxy for constraining the impact of such an effect. We adopt a fiducial value of $M_{\min}=10^{9}M_{\odot}$ for the minimum halo mass in our base model, and study the dependence of the obtained constraints on the assumed value of $M_{\rm min}$ by considering three values for $M_{\min} = \{10^9, 10^{10}, 10^{11}\} M_\odot$. In Figure \ref{fig:Tbar_line}, we show the dependence of the mean brightness temperature $\langle T_{\rm line} \rangle$ on the value of $M_{\rm min}$ in our base models of CO and [CII].

\subsection{Scale-dependent Bias and PNG}\label{ssec:b_NG}
In the presence of PNG, the statistics of initial fluctuations is modified. In the limit of small non-Gaussianity, the dominant NG contribution is from the three-point function or  equivalently its Fourier-space counterpart, i.e. the bispectrum, which is defined as 
\begin{equation}
\langle \zeta_{\vk_1} \zeta_{\vk_2} \zeta_{\vk_3} \rangle = (2\pi)^3 \delta^{(3)} (\vk_{1}+\vk_{2} +\vk_{3})   B_\zeta(k_1, k_2, k_3).
\end{equation}

The local shape PNG \citep{Gangui:1993tt,Wang:1999vf, Verde:1999ij, Komatsu:2001rj} is originated by super-horizon non-linear evolution of curvature perturbations and peaks at squeezed configuration with $k_1 \ll k_2,k_3$. This shape is a probe of multi-field models of inflation, since it is expected to be very small for single-field models \citep{Maldacena:2002vr,Creminelli:2004yq}. In addition to  local shape, we also consider equilateral \citep{Babich:2004gb,Creminelli:2005hu} and orthogonal shapes \cite{Senatore:2009gt}, which can be generated in general single-field models of inflation such as DBI models \citep{Silverstein:2003hf,Alishahiha:2004eh}. Additionally, we consider quasi-single-field model of inflation \citep{Chen:2009zp}, in which the background dynamics is determined by a single degree of freedom but there exist additional scalar fields which can generate non-zero primordial bispectrum. The templates of these shapes are given by
\begin{align}
	B^{\rm loc}_\zeta(k_1, k_2, k_3)  &= \frac{6}{5} f_{\rm NL}^{\rm loc} A_\zeta^2\left[\frac{1}{k_1^{4-n_s} k_2^{4-n_s}}+ 2 \ {\rm perms}\right],  \\
	B^{\rm eq}(k_1,k_2,k_3) &=   \frac{18}{5} A_\zeta^2 f_{\rm NL}^{\rm eq}  \left\{- \frac{2}{(k_1k_2k_3)^{2(4-n_s)/3}} \right. \nonumber \\
	 &\hspace{-.3in} + \left[ -\frac{1}{k_1^{4-n_s} k_2^{4-n_s}} + 2 \ {\rm perms} \right] \nonumber \\
	&\hspace{-.3in}\left. + \left[ \frac{1}{k_1^{(4-n_s)/3}k_2^{2(4-n_s)/3}k_3^{(4-n_s)}} + 5 \ {\rm perms} \right] \right\},  \\
	B^{\rm orth}(k_1,k_2,k_3) &=   \frac{18}{5} A_\zeta^2 f_{\rm NL}^{\rm orth}  \left\{- \frac{8}{(k_1k_2k_3)^{2(4-n_s)/3}} \right. \nonumber \\
	&\hspace{-.3in}+ \left[ -\frac{3}{k_1^{4-n_s} k_2^{4-n_s}} + 2 \ {\rm perms} \right] \nonumber \\
	&\hspace{-.3in}\left. + \left[ \frac{3}{k_1^{(4-n_s)/3}k_2^{2(4-n_s)/3}k_3^{(4-n_s)}} + 5 \ {\rm perms} \right] \right\},  \\	
    B_{\nu}^{\rm qsf}(k_1,k_2,k_3) &=  \frac{54\sqrt{3}}{5}  \frac{A_\zeta^2  f_{\rm NL}^{\rm qsf}}{k_t^{3/2}(k_1k_2k_3)^{3/2} }\frac{Y_{\nu}(8 k_1k_2k_3 /k_t^3)}{Y_{\nu}(8/27)},
\end{align}
where $k_t = \sum_i^3 k_i$, and $Y_{\nu}$ is the Bessel function of the second kind of degree $\nu \equiv \sqrt{9/4-m^2/H^2}$, with $m$ being the mass of the extra scalar field present during inflation.

The total matter fluctuation is related to the curvature perturbations via the Poisson equation. At linear order 
\be
\delta_0(\bk,z) = {\mathcal M}(k,z)\zeta(\bk),
\ee
where 
\be
{\mathcal M}(k,z) = -\frac{2}{5} \frac{k^2 T(k)D(z)}{\Omega_mH_0^2}. 
\ee
$D(z)$ is the linear growth factor normalized to unity today ($D(0) = 1$) and $T(k)$ is the matter linear transfer function that satisfies $T(k\rightarrow 0) =1$. A non-zero primordial bispectrum, therefore, induces non-zero contribution to the matter bispectrum, in addition to that from gravitational evolution. 

In addition to generating a contribution to the matter bispectrum, the presence of PNG leaves an imprint on the power spectrum of biased tracers by inducing a scale-dependent correction to the linear bias. In particular, for halo bias we have
\be\label{eq:scale_dep_bias}
b_h(M,z) \rightarrow \tilde b_h(M,k,z) = b_h(M,z) + \Delta b_h^{\rm NG}(M,k,z).
\ee
For primordial bispectrum of the local shape, the correction to the linear bias has a strong $1/k^2$ scale-dependence \citep{Dalal:2007cu,Matarrese:2008nc,Afshordi:2008ru} 
\be
\Delta b_h^{\rm NG, loc}(M,k,z) =   \frac{6}{5} \frac{f_{\rm NL}^{\rm loc}  \delta_c \left[b_h(M,z)-1\right]}{ {\mathcal M}(k,z)}, 
\ee
which has been used to constrain $f_{\rm NL}^{\rm loc}$ from the power spectrum of LSS biased tracers such as galaxies and quasars \citep{Slosar:2008hx,2013MNRAS.428.1116R,Giannantonio:2013uqa,Giannantonio:2013kqa,2015JCAP...05..040H,Leistedt:2014zqa}. Note that for the local shape, the mass dependence of the scale-dependent bias is the same as the linear Gaussian bias. For a general non-Gaussian shape, this is not necessarily the case. The correction to the halo bias due to a general shape of primordial bispectrum $B_\zeta(k_1,k_2,k_3)$, is given by\footnote{See \cite{Scoccimarro:2011pz} for an alternative derivation.}\citep{Desjacques:2011jb,Desjacques:2011mq}
\begin{align}\label{eq:deltab_NG}
\Delta b_h^{\rm NG}(M,k,z) &=  \frac{2 \mathcal F_R^{(3)}(k,z)}{\mathcal M_R(k,z)}  \\
&\times \left[(b_h(M,z)-1)\delta_c + \frac{d \ln {\mathcal F}^{(3)}_R(k,z)}{d\ln \sigma_R}\right] \nonumber,
 \end{align}
where $\delta_c=1.686$, is the threshold of spherical collapse at redshift zero, and $\sigma_R$ is the variance of the density field smoothed on a scale $R(M) = (3M/4 \pi \bar \rho)^{1/3}$,
\be
\sigma_R^2(z) = \int_0^\infty \frac{dk }{2\pi^2} k^2 P_\zeta(k) \mathcal M_R^2(k,z). 
\ee
$\mathcal M_R(k,z) = W_R(k)\mathcal M(k,z)$, where $W_R(k)$ is the window function which we chose to be Fourier transform of a spherical top-hat filter with radius R,
\be
W_R(k) = \frac{3\left[{\rm sin}(kR) - kR \  {\rm cos} (kR)\right]}{(kR)^3}. 
\ee
The shape factor, ${\mathcal F}^{(3)}_R(k,z)$ is defined as
\begin{align} \label{eq:F3_R}
{\mathcal F}^{(3)}_R(k,z) &= \frac{1}{4 \sigma^2_{R}(z) P_\zeta(k)} \nonumber \\
& \times \int \frac{d^3 q}{(2\pi)^3} \left[\mathcal M_R(q,z) \mathcal M_R(|\bk-\bq|,z) \right. \nonumber \\
&\left. \times B_\zeta(-\bk,\bq, \bk-\bq) \right].
\end{align}
For details of the derivation of the above result, we refer the interested reader to \cite{Desjacques:2011jb,Desjacques:2011mq}. We briefly comment here on the physical interpretation of the two contributions in Equation \eqref{eq:deltab_NG}. The abundance of halos depends on the height of threshold regions, often parametrized as $\nu = \delta_c/\sigma^2_R$.  The presence of primordial bispectrum, modulates the variance of the small scale density field such that the variance and hence $\nu$, vary from point to point. This is the origin of the first term in square brackets in Equation \eqref{eq:deltab_NG}. Additionally, variation of $\nu$ also changes the mapping between mass interval and threshold height, which gives rise to the second term in Equation \eqref{eq:deltab_NG}. This mass-dependent contribution is important for most shapes of primordial bispectrum apart from the local shape, in particular for lower mass halos. As is shown for in Figs. \ref{fig:L_CO} and \ref{fig:L_CII}, at $z=2$ for instance, halos of masses of $\sim 10^{12} M_\sun h^{-1}$ are the dominant contribution to the CO/CII luminosities. We therefore use the full expression given in Equation \eqref{eq:deltab_NG} in our analysis.

As emission lines, such as that from CO and [CII], are biased tracers of the underlying dark matter distribution, the scale-dependent correction to their bias can be used to obtain constraints on the level of PNG (e.g., \citet{Camera:2013kpa,Camera:2014bwa,Alonso:2015uua}). In the presence of non-zero primordial bispectrum, the line bias is given by Equation \eqref{eq:line_bias}, replacing the Gaussian linear halo bias $b_h(M,z)$ by $\tilde b_h(M,k,z)$, as defined in Equation \eqref{eq:scale_dep_bias}, with non-Gaussian correction given by Equation \eqref{eq:deltab_NG}.

\begin{figure}\centering
\includegraphics[width=0.5 \textwidth]{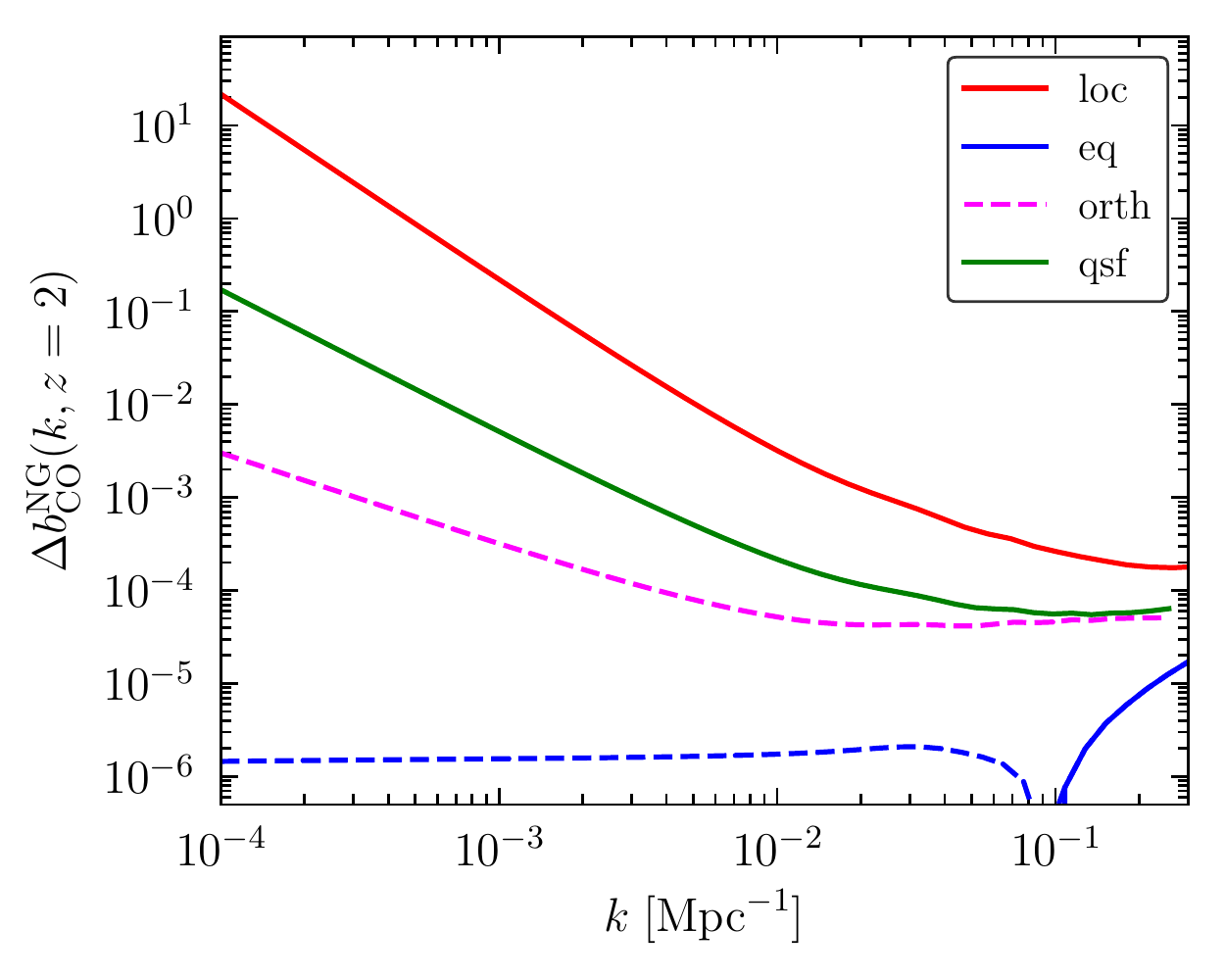}
\includegraphics[width=0.5 \textwidth]{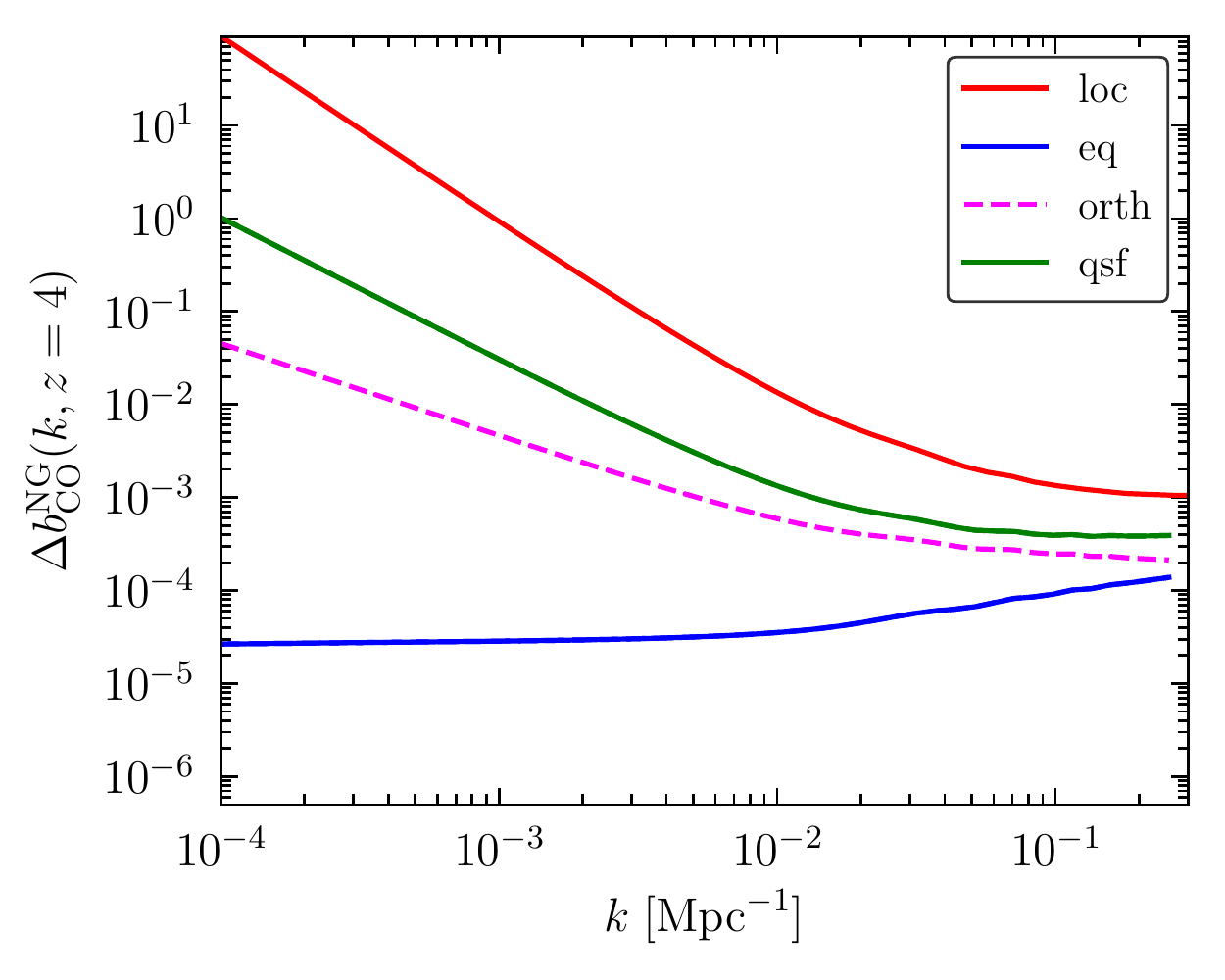}
\caption{The correction to linear bias of CO due to primordial bispectrum at $z=2$ (top) and $z=4$ (bottom). Dashed (solid) lines indicate negative (positive) values. We considered $f_{\rm NL}^{\rm shape} = 1$ and for qsf model, $\nu=1$.}
\label{fig:deltab}
\end{figure}

In Fig. \ref{fig:deltab} we show the correction to the CO bias due to PNG of local, equilateral, orthogonal and quasi-single-field models, given in Equation \eqref{eq:deltab_NG} for $z=2$ ($z=4$) in the top (bottom) panel. The dashed lines indicate negative values. Note that at a fixed redshift, the lower the mass of halos, the larger is the contribution from the second term in Eq.\eqref{eq:deltab_NG}. For a fixed halo mass, the higher the redshift, the less significant is the second terms. 

Note that in the presence of PNG, there are two additional contributions to the halo bias \citep{Desjacques:2008vf} that we have neglected, first is a scale-independent contribution which arises due to the modification of the halo mass function due to PNG \citep{Slosar:2008hx,Afshordi:2008ru}. The second is a scale-dependent correction due to corrections to matter power spectrum in the presence of PNG \citep{Scoccimarro:2003wn,Taruya:2008pg,Pillepich:2008ka}. We also neglect the corrections to the line bias that are directly induced by the modification of the mass function \citep{LoVerde:2007ri} in Equation \eqref{eq:line_bias}. 

\subsection{Additional Effects}\label{ssec:RSD_AP}
We account for two additional effects in modeling the observed power spectrum: redshift-space distortion, and the Alcock-Paczynski effect. RSD arises from the fact that the power spectrum is measured in redshift space rather than real-space, where the peculiar velocities of galaxies modify their distribution. At linear level, the effect is captured by Kaiser effect \citep{Kaiser:1987qv} which enhances the matter density field by a factor of $(1+f\mu^2)$, where $f$ is the logarithmic growth factor. At non-linear level, its effect is parametrized as a suppression factor on small-scale densities. The line power spectrum therefore becomes
\begin{align}
P_{\rm clust}^s(k,\mu,z) &= P_{\rm clust}(k,z)\left[1+\mu^2\beta(k,z)\right]^2 \nonumber \\
&\times {\rm exp}\left(-\frac{k^2 \mu^2 \sigma_v^2}{H^2(z)}\right),
\end{align}
where $\beta(k,z) = f/b_{\rm line}(k,z)$, $\sigma_v$ is the 1D pairwise velocity dispersion of the galaxies, $H(z)$ is the Hubble expansion rate and $P_{\rm line}(k,z)$ is given by Equation \eqref{eq:ps_r} replacing $b_1(z)$ by $\tilde b_1(k,z)$ as in Equation \eqref{eq:scale_dep_bias}. The velocity dispersion is further defined as
\be
\sigma_v^2 = (1+z)^2\left[\frac{\sigma_{\rm FOG}^2(z)}{2} + c^2 \sigma_z^2\right], 
\ee
and has two contributions at a given redshift: one from the finger-of-god (FOG) effect \citep{Seo:2007ns} and the other due to the intrinsic line width of individual emitters, over which the emission is smeared out. We assume the redshift dependence for the FOG effect to be
\be
\sigma_{\rm FOG}(z) = \sigma_{{\rm FOG},0}\sqrt{1+z},
\ee
and vary a $ \sigma_{{\rm FOG},0}$ as a free parameter in our forecast. We further assume $\sigma_z = 0.001 (1+z)$.

The second effect, the Alcock-Pazcynski effect originates from the fact that we infer the distances from the observed redshifts and angular position of galaxies assuming a reference cosmology. If the reference cosmology is different than the ``true'' one, the radial and transverse distances are distorted with respect to the true values. The observed line intensity power spectrum of the reference cosmology is related to that in true cosmology as \citep{Seo:2003pu}
\be
{\tilde P}_{\rm clust}^s(\tilde k,\tilde \mu,z) = \frac{H_{\rm true}(z)}{H_{\rm ref}(z)}  \left[\frac{D_{A,{\rm ref}}(z)}{D_{A,{\rm true}}(z)}\right]^2 P_{\rm clust}^s(k,\mu,z),
\ee
where $D_A$ is the angular diameter distance and the distance ratios account for the difference in volume between the two cosmologies. The tilde coordinates are those in reference cosmology and are related to those in true cosmology as 
\begin{align}
k &= \tilde k \left[(1-\tilde \mu^2) \frac{D_{A,{\rm ref}}^2(z)}{D_{A,{\rm true}}^2(z)} + \tilde \mu^2 \frac{H_{\rm true}^2(z)}{H_{\rm ref}^2(z)}\right]^{1/2},  \nonumber \\
\mu &= \tilde \mu \frac{\tilde k}{k}\frac{H_{\rm true}(z)}{H_{\rm ref}(z)}.
\end{align}
We take the reference cosmology to be our fiducial model, as will be discussed in Section \ref{sec:forecast}. 

\section{Instrument and Survey Description}\label{sec:survey_design}
For our analysis, we evaluate a given survey as a function of five different primary attributes: the per-mode noise, the sky coverage, the redshift coverage, as well as the minimum and maximum spatial scales recoverable within the survey.
\subsection{Instrumental Noise}\label{ssec:inst_noise}
For a single mode, $\textbf{k}$, the estimated instrumental noise is given by
\begin{equation}\label{eq:inst_noise_power}
P_{\textrm{N}} = \sigma_{\rm N}^{2}V_{\textrm{vox}},
\end{equation}
where $\sigma_{\rm N}$ is the RMS noise within a single voxel, of volume $V_{\textrm{vox}}$, in our hypothetical image cube. This volume can be further defined as
\begin{equation}\label{eq:v_vox}
V_{\rm vox} = \Omega_{\rm b}\delta\nu \left (\frac{dl}{d\theta} \right)^{2}  \frac{dl}{d\nu},
\end{equation}
where $\Omega_{\rm b}$ is the field-of-view for the instrument, $\delta\nu$ is the width (resolution) of a single frequency channel, and the terms $dl/d\nu$ and $dl/d\theta$ reflect the change in units. Utilizing the radiometer equation, we can rewrite $\sigma_{\rm N}$ as
\begin{equation}\label{eq:radiometer}
\sigma_{\rm N} = \frac{T_{\textrm{sys,RJ}}}{\tau_{\textrm{int}}\delta\nu \sqrt{N_{\rm pol}N_{\textrm{beam}}}},
\end{equation}
where $T_{\textrm{sys,RJ}}$ is the system temperature of the instrument (under the Rayleigh-Jeans approximation), $\tau_{\textrm{int}}$ is the integration time per single pointing of the instrument, $N_{\rm pol}$ is the number of polarizations the instrument is sensitive to, and $N_{\textrm{beam}}$ is the number of independent spatial positions (i.e., beams) instantaneously sampled on the sky. We note here that for incoherent detectors -- like those typically used in the sub-mm/FIR regime that the [CII] line will be found in -- one typically reports the sensitivity in terms of noise-equivalent power (NEP; typically expressed in units of W Hz$^{-1/2}$). The NEP, defined in terms of the power absorbed by the detector (i.e., the electrical NEP), can be related to $T_{\rm sys,RJ}$ as
\begin{equation}
T_{\rm sys,RJ} = \frac{\textrm{NEP}}{k_{\rm B} \eta_{\rm inst}\sqrt{2\delta\nu}},
\end{equation}
where $\eta_{\rm inst}$ is the end-to-end optical efficiency (including the optical throughput and absorber efficiency of the instrument). Assuming total integration time, $\tau_{\textrm{tot}}$,{} is equal to $\tau_{\rm int}$ multiplied by the ratio of the total survey area, $\Omega_{\textrm{survey}}$, to that of the primary beam (i.e., the number of independent pointings required for a given survey), and combining Equations \eqref{eq:inst_noise_power} and \eqref{eq:radiometer}, we find
\begin{equation}\label{eq:inst_noise_power_ext}
P_{\rm N} = \frac{T_{\rm sys}^2}{\tau_{\rm tot}N_{\rm beam}}\Omega_{\rm surv}\left(\frac{dl}{d\theta}\right)^2 \frac{dl}{d\nu}.
\end{equation}
In Equation \eqref{eq:inst_noise_power_ext}, we assume that the per-mode noise is primarily dominated by the thermal or photon noise of the instrument, absent significant contributions from continuum or other foreground (or background) contaminants that may affect a given survey. We further discuss the validity of this assumption, and what how such contamination may limit a given survey, in Section \ref{ssec:disc_max_scale}.
\subsection{Redshift Coverage}\label{ssec:surv_red}
In principle, surveys targeting higher redshift provide larger volumes, giving access to more modes on the large scales where the constraining power on $f_{\rm NL}^{\rm loc}$ is the greatest. However, this advantage can be offset by several other aspects, not the least of which is that the emission from high redshift may be much weaker than that from lower redshift (as shown in Figure \ref{fig:Tbar_line}). The situation is further complicated by he frequency-dependent contributions to the instrumental noise (e.g., limited transmission of the atmosphere at high frequencies), which may further drive one to a particular redshift range. We discuss these effects further in Section \ref{ssec:instrument_types}, and note that in our analysis, we assume no meaningful emission exists beyond $z=10$, such that there is no direct gain for a given survey to probe beyond this redshift.
\subsection{Sky Coverage}\label{ssec:surv_fsky}
Similar to the redshift coverage, access to larger areas of the sky affords a given survey the ability to probe larger volumes, and therefore larger modes. However, this increase in power is offset by decreased time spent per position on the sky, translating to increased noise per-voxel and per-mode, as shown in Equation \eqref{eq:inst_noise_power}. Additionally, accessing larger areas of the sky may necessitate observing in areas of the sky where Galactic foreground emission becomes increasingly significant. The ability to access these areas depends not only on the ability to accurately model such emission, but also to model and well-calibrate the instrumental response. While we have not explicitly modeled this emission in our analysis, we assume that this emission is likely to produce correlated errors in the northern and southern Galactic hemispheres, such that the largest mode accessible is set (in part) by the largest contiguous patch of sky observed in a survey.

For the purposes of our analysis, we set the maximum allowable value of $f_{\rm sky}$ to be 80\%, assuming that at least 20\% will be sufficiently contaminated by Galactic emission to make intensity mapping analyses of this area impractical. This limit is likely an optimistic upper threshold as to what could be achieved with a large-area mm/sub-mm survey \citep{Switzer:2017kkz} -- in Section \ref{ssec:results_surv_params}, we explore further how the change of $f_{\rm sky}$ impacts our final estimate for $\sigma(f_{\rm NL})$ in a given survey.
\subsection{Maximum Recoverable Length Scale}\label{ssec:surv_kmin}
While the total volume -- set by the redshift and the sky coverage -- sets the maximum length-scale theoretically accessible by a given survey, foreground emission and errors in instrument calibration will set limits on which modes are practical for use in an intensity mapping analysis. We describe the lowest wavenumber recoverable as $k_{\rm min}$, where $k_{{\rm min},\perp}$ and $k_{{\rm min},\parallel}$ are the minimum wavenumbers in the direction perpendicular and parallel to the line of sight. The value of $k_{\rm min}$ is set in part by two distinct groups of modes that are likely to be corrupted by foregrounds, which we will want to exclude or otherwise down-weight in our analysis. 

The first set of modes are those purely perpendicular to the line of sight, where the power from continuum emission is likely to vastly exceed that from the target of interest \citep{2015ApJ...814..140K}. Even with accurate models, non-smooth spectral structure can be imparted by the frequency-dependent response of the instrument, making a sheet of modes around $k_{z}=0$ difficult to utilize in an intensity mapping auto-correlation measurement. As the bandwidth of a given redshift interval grows, so too does the sum total continuum power within the map, the residual power from which will eventually bleed into modes outside of the $k_{z}=0$ sheet, which can add systematic noise to our measurement. Rather then calculating this contribution -- which depends heavily on the instrument respones -- we instead define $\eta_{\rm min}$ as the ratio of the observed frequency divided by the maximum bandwidth over which the frequency-dependent response of the instrument (in the presence of foregrounds) is assumed to be smooth. We  then define the maximum recoverable spatial scale in the direction parallel to the line of sight as 
\begin{equation}\label{eq:kpar_min}
k_{\parallel,{\rm min}} = 2\pi \eta_{\rm min} \left [ \nu_{obs} \frac{dl}{d\nu} \right ] ^{-1}.
\end{equation}
As a practical consideration, we note here that the lowest value of $\eta_{\rm min}$ possible for a given experiment is set by the redshift coverage of that experiment (e.g., for a survey where $\Delta z/z = 0.1$, $\eta_{\rm min} \geq 10$). We further impose a limit of $\eta_{\rm min} \geq 0.5$. We also note that under this framework, there may be indirect gain from measurements outside the nominal redshift window of $z=[0-10]$ discussed in Section \ref{ssec:surv_red}, as the increased bandwidths \textit{may} allow one to achieve lower values of $\eta_{\rm min}$ inside the frequency range of interest.

The second set of likely corrupted modes are those purely parallel to the line of sight, which can be affected by position-dependent errors in the estimated total power (i.e., frequency-independent amplitude errors), arising from either instrumental noise or foregrounds. In the 3D power spectrum, this will make a cylinder of modes around the $k_{z}$ axis relatively difficult for an intensity mapping experiment to access, where the width of this cylinder is set by the angular diameter of the survey area, which can be written as a function of the sky-coverage for the experiment. For the purposes of our analysis, we assume this limit is set by the survey area of a single contiguous patch of the sky, and therefore the maximum recoverable spatial scale in the direction perpendicular to the line of sight is given by
\begin{equation}\label{eq:kperp_min}
k_{\perp,{\rm min}}\approx2\pi \left[2\sin\left (\theta_{\rm max}/2\right )\frac{dl}{d\theta} \right ]^{-1},
\end{equation}
where $\theta_{\rm max}$ is the largest angular scale covered by a given survey. Generally speaking, we assume that $\theta_{\rm max}=\sqrt{\Omega_{\rm surv}}$, although we note that for very large-area surveys, $\theta_{\rm max}$ may be limited by Galactic emission. Given that angular scales larger than 180 degrees may suffer from degraded sensitivity due to increased foregrounds from Galactic emission, we set a limit of $\theta_{\rm max}<160^{\circ}$ (i.e., surveys are restricted to $\beta>10^{\circ}$). We further assume that for $f_{\rm sky}>0.4$, area coverage is equally split in half between northern and southern Galactic hemispheres, such that $\theta_{\rm max}=\sqrt{\Omega_{\rm surv}/2}$. This effectively down-weights the largest angular scales a full-sky survey hypothetically has access to, which are most likely to be contaminated by Galactic foreground emission \citep{Switzer:2017kkz}. We note that at high redshift, this has only a marginal impact on the lowest value of $k$ that can be accessed by a given survey, which is generally more sensitive to the value of $\eta_{\rm min}$.

To account for the loss of the aforementioned modes, we assume that there is some scale over which the combination of astrophysical foregrounds (or their residuals after modeling) and the instrumental response are relatively smooth, which can then be subtracted out of the data. This is equivalent to dropping modes where $k_{z}=0$ or $k_{x}=k_{y}=0$, with some loss of sensitivity in modes around $k_{z}\approx0$ or $k_{x}\approx k_{y}\approx0$, the degree of which depends on the size and shape of the smoothing kernel. Assuming two kernels are used (one for the spatial domain, one for the spectral), the size of which are set by $k_{\rm min,\perp}$ and $k_{\rm min,\parallel}$, then we estimate the attenuation of the signal of interest as
\begin{align}\label{eqn:atten_kmin}
\alpha_{\rm min}(k_{\perp},k_{\parallel}) &= \left (1-e^{-k_{\perp}^{2}/(k_{\perp,{\rm min}}/2)^{2}}\right)  \nonumber \\
&\times \left (1-e^{-k_{\parallel}^{2}/(k_{\parallel,{\rm min}}/2)^{2}} \right)
\end{align}
We note that the loss estimated in Equation \eqref{eqn:atten_kmin} is only an approximation, and does not account for calibration errors that may cast power into modes at higher values of $k$. However, this additional source of error can be, to first-order, approximated as an increase in the per-mode noise in the power spectrum. While not explicitly modeled in our analysis, we discuss a related scenario under which the effective integration time is reduced (leading to an increase in the per-mode noise) in Section \ref{ssec:results_surv_params}. We further note that $\eta_{\rm min}$ and $\theta_{\rm max}$ may reflect optimistic estimates for a given survey, and further consider the impact of more pessimistic limits on our optimized survey in Section \ref{ssec:disc_max_scale}.
\subsection{Minimum Recoverable Length Scale}\label{ssec:surv_kmax}
The minimum size-scale recovered, which corresponds to the maximum wavenumber $k_{\rm max}$ at which our instrument has sensitivity, is set by the spatial and spectral resolution of the instrument. Larger apertures (with smaller fields of view) and finer frequency resolution correspond to greater values $k_{\rm max}$, while coarser spectral resolution and smaller aperture sizes correspond to smaller values. We can further define the minimum scale recoverable in the direction perpendicular to the line of sight as
\begin{equation}\label{eqn:kmin_perp}
k_{{\rm max},\perp} \approx 2\pi \left [ \frac{c/\nu_{obs}}{D_{\rm ant}} \frac{dl}{d\theta}\right ]^{-1},
\end{equation}
and the minimum scale recoverable in the direction parallel to the line of sight as
\begin{equation}\label{eqn:kmin_par}
k_{{\rm max},\parallel} \approx  2\pi \left [ \delta\nu  \frac{dl}{d\nu}\right ] ^{-1}.
\end{equation}
In Equation \eqref{eqn:kmin_perp}, $D_{\rm ant}$ is the diameter of the aperture used for our hypothetical survey. 
We assume the attenuation of the signal due to finite resolution of th instrument to be
\begin{equation}\label{eqn:atten_kmax}
\alpha_{\rm max}(k_{\perp},k_{\parallel}) = e^{-(k_{\perp}{^2}/k_{\perp,{\rm max}}^{2}+k_{\parallel}{^2}/k_{\parallel,{\rm max}}^{2})}
\end{equation}

For local shape PNG, the constraining power for a given survey is expected to be relatively insensitive to the value of $k_{\rm max}$, as the enhancement in power is strongest on the largest length scales (i.e., modes with wavenumbers $k<10^{-2}$). Our final estimate for the constraining power for a given survey is relatively insensitive to the value of $k_{\rm max}$. However, for other shapes, the constraints on PNG can significantly depend on the value of $k_{\rm max}$. 
\subsection{Generalized Instrument Description}\label{ssec:instrument_types}
While there are existing and upcoming instruments for conducing CO and [CII] intensity mapping-focused surveys, none exist that are specifically focused on probing PNG. Therefore, rather than focusing on specific existing instruments, as was done in \citet{MoradinezhadDizgah:2018zrs}, we seek to define the scope of the instrument required to achieve $\sigma(f_{\rm NL}^{\rm loc})=1$, while minimizing the effective ``cost'' of such an instrument. As a simple, first-order approximation, we assume that this cost is proportional to 
\begin{equation}\label{eq:inst_cost}
\chi_{\rm inst} = N_{\rm beam}\log_{2}(\nu_{\rm max}/\nu_{\rm min})\tau_{\rm int,hr},
\end{equation}
where $\tau_{\rm int,hr}$ is the total on-source integration time for a given survey (in hours), while $\nu_{\rm min}$ and $\nu_{\rm max}$ are the minimum and maximum frequency covered by a given survey. We further define the number of octaves of spectral coverage as $N_{\rm oct}=\log_{2}(\nu_{\rm max}/\nu_{\rm min})$. As it is constructed, Equation \eqref{eq:inst_cost} produces a value with units of beam-octave-hours (abbreviated henceforth as BOHs). For both CO and CII, we consider an instrument of fixed aperture size, spectral resolution, and given noise performance, allowing the sky and redshift coverage to vary as free parameters.

As the atmosphere is generally transparent below 50 GHz, a ground-based instrument would be capable of surveying CO(1-0) at $z\gtrsim1$. For this instrument, we assume a 6-meter aperture with spectral resolving power, $R=\nu_{obs}/\delta\nu$, of $R=1000$ (i.e., 300 km/sec spectral resolution). For modeling the system temperature, we utilize a simple model based on the performance of present and future heterodyne-based radio astronomy instruments for centimeter wavelengths (e.g., \citealt{Velazco:2017,doi:10.1117/12.670472,2018IAUS..336..426M}), where 
\begin{equation*}\label{eqn:tsys_CO}
T_{\rm sys} = 
\begin{cases}
\ \nu_{obs}\ (\textrm{K}/\textrm{GHz})\ & : \nu_{obs}\geq20\ \textrm{GHz}, \\
\ 20\ \textrm{K} & : \nu_{obs} < 20 \textrm{ GHz}.
\end{cases}
\end{equation*}
As the experiment under consideration is ground-based, we assume that only one of two Galactic hemispheres is sufficiently accessible for our hypothetical CO intensity mapping survey, such that $f_{\rm sky}<0.4$.

In contrast to CO(1-0), the 158-micron line of [CII] is relatively inaccessible from the ground at $z\lesssim4$, due to limited atmospheric transmission at sub-mm wavelengths. We therefore consider a space-based experiment for [CII] intensity mapping, modeled loosely on the \textit{Planck} satellite mission. For this instrument, we consider a 1.5-meter aperture with $R=80$, and a passively cooled aperture with a temperature of 40 K and 1\% emissivity, in rough agreement with the design specifications for \textit{Planck} \citep{Planck2011:ER02}. In modeling the sensitivity of the instrument, we include photon noise contributions of the CMB ($T_{\rm B}=2.725$ K), as well as that of Galactic ($T_{\rm B}=18$ K) and Zodiacal dust ($T_{\rm B}=240$ K). We assume that the emissivity of the dust ($\varepsilon$) can be described as a function of frequency,
\begin{equation}
\varepsilon = \varepsilon_{\circ}(\nu/\nu_{\circ})^{\beta},
\end{equation}
where $\varepsilon_{\circ}=2\times10^{-4}$, $\nu_{\circ}=3$ THz, $\beta=2$ for Galactic dust emission, and $\varepsilon_{\circ}=3\times10{^-7}$, $\nu_{\circ}=2$ THz, $\beta=2$ for Zodiacal dust emission \citep{Finkbeiner:1999gal,Fixsen:2002zod}. We assume our [CII]-focused instrument is background photon noise limited, requiring a detector NEP of $\lesssim 10^{18}\ \textrm{W Hz}^{1/2}$, in line with an upcoming generation of advance detectors for FIR astronomy (e.g., \citealt{Khosropanah:2016nep}).

\section{Forecasting methodology}\label{sec:forecast}
\subsection{Fisher Information Matrix}
We use the Fisher matrix formalism to forecast the potential of future CO and [CII] intensity mapping surveys to constrain PNG of several shapes, namely local, equilateral, orthogonal and quasi-single-field, focusing primarily on the local shape. In general, the Fisher information matrix is defined as 
\be
F_{\alpha \beta} = - \left< \frac{\partial^2 {\rm ln} \ {\mathcal L}({\bf x}, \boldsymbol{\lambda})}{\partial \lambda_\alpha \partial \lambda_\beta}\right>,
\ee
where ${\mathcal L}$ is the likelihood of the data ${\bf x}$ given the parameters $\boldsymbol {\lambda}$. In the case that all parameters are fixed except for say $\alpha_{\rm th}$ parameter, the 1$\sigma$ error on this parameter is $\sigma (\lambda_{\alpha}) = 1/\sqrt{F_{\alpha \alpha}}$. Otherwise, if marginalized over all the other parameters, $\sigma (\lambda_{\alpha}) = \sqrt{F_{\alpha \alpha}^{-1}}$. 

Assuming a Gaussian likelihood, for a data of mean $\mu = <{\bf x}>$ and covariance matrix ${\bf C} \equiv \langle {\bf x}{\bf x}^T\rangle  -{\boldsymbol \mu}{\boldsymbol \mu}^T$, the Fisher matrix can be written as:
\be \label{eq:gen_Fish}
F_{\alpha \beta} = \frac{1}{2}{\it tr} \left[{\bf C}_{,\alpha}{\bf C}^{-1}{\bf C}_{,\beta}{\bf C}^{-1}\right] + {\boldsymbol \mu}_{,\alpha}^T {\bf C}^{-1}{\boldsymbol \mu}_{,\beta},
\ee
where $(_{,\alpha})$ denotes the derivative with respect to parameter $\alpha$. For the 3D intensity mapping,  the average power spectrum in a thin shell of radius $k_i$, width $dk_i$ and volume $V_i = 4\pi k_i^2 dk_i/(2\pi)^3$ in Fourier space is measured. Therefore, for each redshift bin and angle, we have a non-zero mean and covariance. The dominant term of the Fisher matrix is, thus, the second term in Equation \eqref{eq:gen_Fish}. 

For a redshift bin $z_i$, the Fisher matrix is given by:
\begin{align}\label{eq:Fisher_single}
F_{\alpha\beta}^i &=  \int_{-1}^1\int_{k_{\rm min}}^ {k_{\rm max}}  \frac{ k^2 {\rm d}k \ {\rm d}\mu  }{8 \pi^2} \ V_{\rm eff}(k,\mu,z_i) \nonumber \\
&\times \frac{\partial {\rm ln} {\tilde P}^s_{\rm clust}(k,\mu,z_i)}{\partial { \lambda}_\alpha } \frac{\partial {\rm ln} {\tilde P}^s_{\rm clust}(k,\mu,z_i)}{\partial {\lambda}_\beta } ,
\end{align}
where $V_{\rm eff}$ is the effective volume of a given redshift bin defined by:
\be
V_{\rm eff} \simeq   \left[\frac{\tilde P_{\rm clust}^s(k,\mu,z_i)}{\tilde P_{\rm clust}^s(k,\mu,z_i) + P_{\rm shot}(z_i) + \tilde P_N(k,\mu)} \right]^2 V_i,
 \label{eq:vol_eff}
\ee
where $\tilde P_{\rm clust}^s$ is the clustering component of CO or [CII] power spectrum in redshift-space, $P_{\rm shot}$ is the shot-noise given in Equation \eqref{eq:ps_shot} and $\tilde P_{\rm N}(k,\mu,z) = P_{\rm N} \alpha^{-1} _{\rm max}(k,\mu)\alpha^{-1} _{\rm min}(k,\mu) $ is the effective instrumental noise as discussed in Section \ref{sec:survey_design}. For a survey covering a fraction of the sky $f_{\rm sky}$, the volume of a redshift bin between $z_{\rm min}$ and $z_{\rm max}$ is: 
\be
V_i=\frac{4\pi}{3} \times f_{\rm sky} \left[d_c^3(z_{\rm max}) -d_c^3(z_{\rm min}) \right],
\ee
where $d_c(z)$ is the comoving distance from redshift $z$ 
\be
d_c(z) = \int_0^z \frac{c}{H(z)} dz.
\ee

In all our forecasts, we vary 5 cosmological parameters: the amplitude $A_s$, and the spectral index $n_s$ of the primordial fluctuations, the Hubble parameter $h$, the energy density of cold dark matter $\Omega_{\rm cdm}$ and baryons $\Omega_b$. Additionally we vary the amplitude of primordial bispectrum of a given shape $f_{\rm NL}^{\rm shape}$, as well as a single parameter $\sigma_{\rm FOG0}$ for the velocity dispersion (assuming that the redshift-dependence of the velocity dispersion is given by $\sigma_{\rm FOG} = \sigma_{\rm FOG0} \sqrt{z}$). When considering the PNG shape from quasi-single model, we also vary the parameter $\nu$, which quantifies the mass of the extra scalar field present during inflation.  In our base model, we fix the line bias to that given in Equation \eqref{eq:line_bias}. We also consider the case that no prior knowledge of the the values of the biases is assumed and biases in each redshift bin are taken as a free parameter to be marginalized over. Therefore, when varying the biases, our parameter array is given by: ${\boldsymbol\lambda} = \left[{\rm ln} (10^{10}A_s), n_s, h,\Omega_{\rm cdm},\Omega_b, f_{\rm NL}^{\rm shape}, \sigma_{{\rm FOG},0}, {\bf b} \right]$, where $\bf b$ is an array of biases in the redshift bins considered. We set the fiducial values of the cosmological parameters to that from Planck 2015 data \citep{2016A&A...594A..13P}  with ${\rm ln} (10^{10}A_s) = 3.067, n_s = 0.9672, h = 0.6778,\Omega_{\rm cdm} =0.2583,  \Omega_b = 0.04856$. We choose the fiducial value of $f_{\rm NL}^{\rm shape}=1$, $\nu = 1$ and $\sigma_{{\rm FOG},0} = \ 250 \ {\rm km} s^{-1}$. When varying the biases, their fiducial values are set according to Equation \eqref{eq:line_bias}.

For each redshift bin, we take  $k_{\rm min} = 2 \pi(3V_i/4\pi)^{-1/3}$ where $V_i$ is the volume of the corresponding bin. We set the $k_{\rm max} = 0.15$ h ${\rm Mpc}^{-1}$ at $z=0$. At higher redshifts, we obtain the $k_{\rm max}$ such that the following condition for the variance of the linear matter density field is satisfied:
\be
\sigma^2(z) = \int_{k_{\rm min}(z)}^{k_{\rm max}(z)} \frac{d^3 k}{(2\pi)^3} P_0(k,z) = {\rm const} = \sigma^2(z=0).
\ee 
This choice ensures that at a given redshift $z$, we are in the regime where the fluctuations in matter density are in nearly linear regime where the perturbation theory is valid. We further impose a more conservative upper bound of $k_{\rm max} \leq 0.3$ to ensure that our assumptions of linear bias and linear RSD (linear Kaiser term) are valid.   

\section{Results}\label{sec:results}
\begin{figure}[!t]
\begin{center}
\includegraphics[scale=.58]{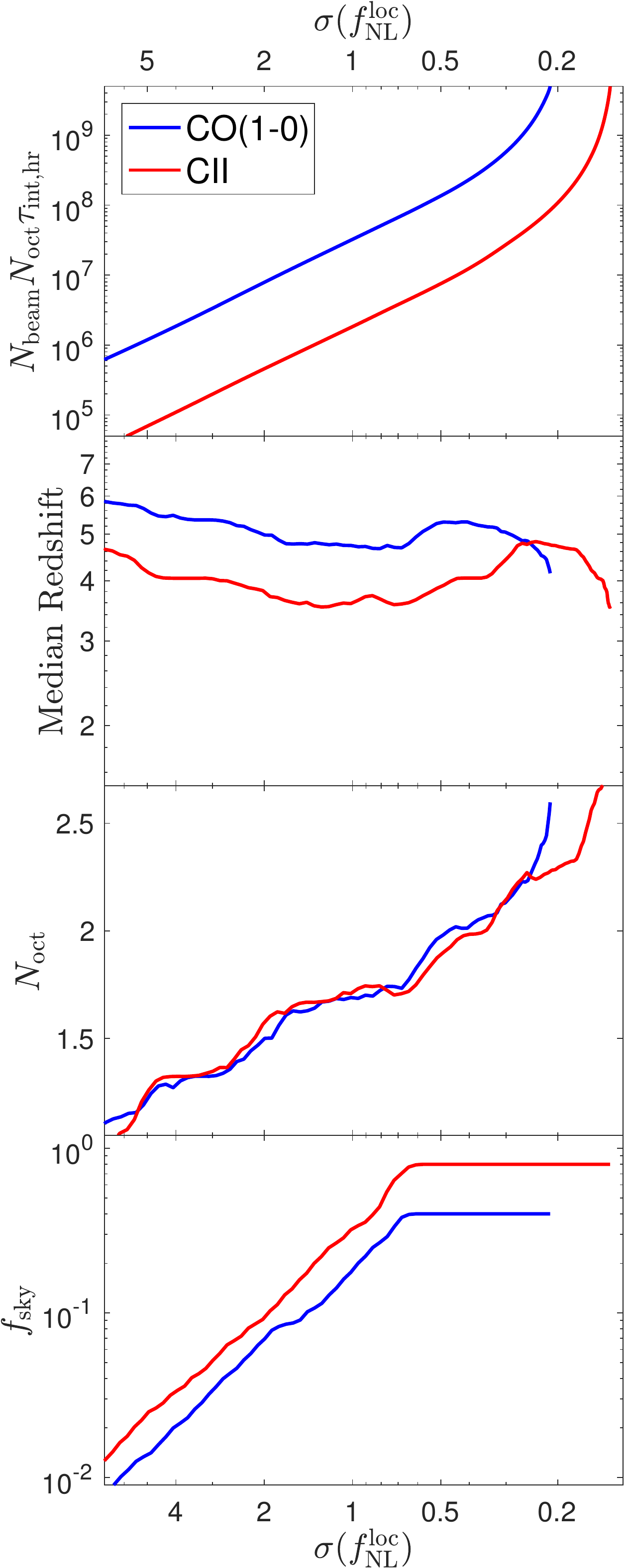}
\caption{The minimum number of BOHs required to achieve a target $\sigma(f_{\rm NL}^{\rm loc})$ is shown (top panel), with the median redshift (second panel), the spectral coverage expressed as the number of octaves (third panel), and the required sky coverage (bottom panel) for an optimized survey to achieve said target. The survey constraints and instrument noise properties are given in Section \ref{ssec:instrument_types}. In the top panel, the best achievable values for $\sigma(f_{\rm NL}^{\rm loc})$ differ for CO and [CII] primarily due to the difference in maximum sky coverage: our ground-based setup for CO is assumed to have access to half of the sky (and thus, approximately half the number of unique modes) available to the space-based setup for [CII].
\label{fig:optimal_instrument}}
\end{center}
\end{figure}
For our survey optimization analysis, we evaluate the best achievable value of $\sigma(f_{\rm NL}^{\rm loc})$ given a fixed cost $\chi_{\rm inst}$ by using a simulated annealing algorithm to identify the optimal sky and redshift coverage. The results of this analysis are shown in Figure \ref{fig:optimal_instrument}, where we have imposed the Planck priors on cosmological parameters. Given the aforementioned survey constraints found in Section \ref{ssec:instrument_types}, we find that for CO, achieving $\sigma(f_{\rm NL}^{\rm loc})=1$ requires a minimum of $3.2 \times 10^{7}$ BOHs, which optimally requires redshift coverage of $z=[2.2-9.3]$ and sky coverage of $f_{\rm sky}=0.18$. For [CII], achieving the same target for $\sigma(f_{\rm NL}^{\rm loc})$ requires a minimum of $1.8 \times 10^{6}$ BOHs, with redshift coverage of $z=[1.5-7.2]$, and sky coverage of $f_{\rm sky}=0.34$. 

In the top row of Figure \ref{fig:inst_params}, we show how adjusting each of the survey parameters (fixing all the others) affects the constraint on $\sigma(f_{\rm NL}^{\rm loc})$ for both CO and CII intensity mapping surveys, compared to optimized survey parameter values. We find that the optimized surveys are only moderately sensitive to the input parameters, with changes of 10\% (or more, in the case of $N_{\rm oct}$ and $f_{\rm sky}$ producing negligible changes ($<5\%$) in $\sigma(f_{\rm NL}^{\rm loc})$. In the matrix plots shown at the bottom of Figure \ref{fig:inst_params}, we show the impact of varying simultaneously two of the four survey parameters for CO (shown in plots above the diagonal) and CII (plots below the diagonal). The contours correspond to constant $\sigma(f_{\rm NL}^{\rm loc})$ constraints with values indicated on the lines.

\subsection{Dependence on Survey Parameters}\label{ssec:results_surv_params}
\noindent\emph{Effective integration time requirements:} Shown in the top panel of Figure \ref{fig:optimal_instrument} is the best value of $\sigma(f_{\rm NL}^{\rm loc})$ achievable (with constraints on detector noise and survey parameters) as a function of BOHs -- a proxy for the effective integration time (i.e., the number of detectors multiplied by the integration time). For both CO and [CII] experimental setups, the constraints improve as $\chi^{-1/2}$, until one reaches $\sigma(f_{\rm NL}^{\rm loc})\approx 0.4$, at which point added integration time provides diminishing returns. Not coincidently, this is also the point at which the volume of the surveys are roughly maximized, such that cosmic variance contributes an increasingly large fraction to the overall uncertainty in the constraint.

\begin{figure*}[!h]
\begin{center}
\includegraphics[width=0.95\textwidth]{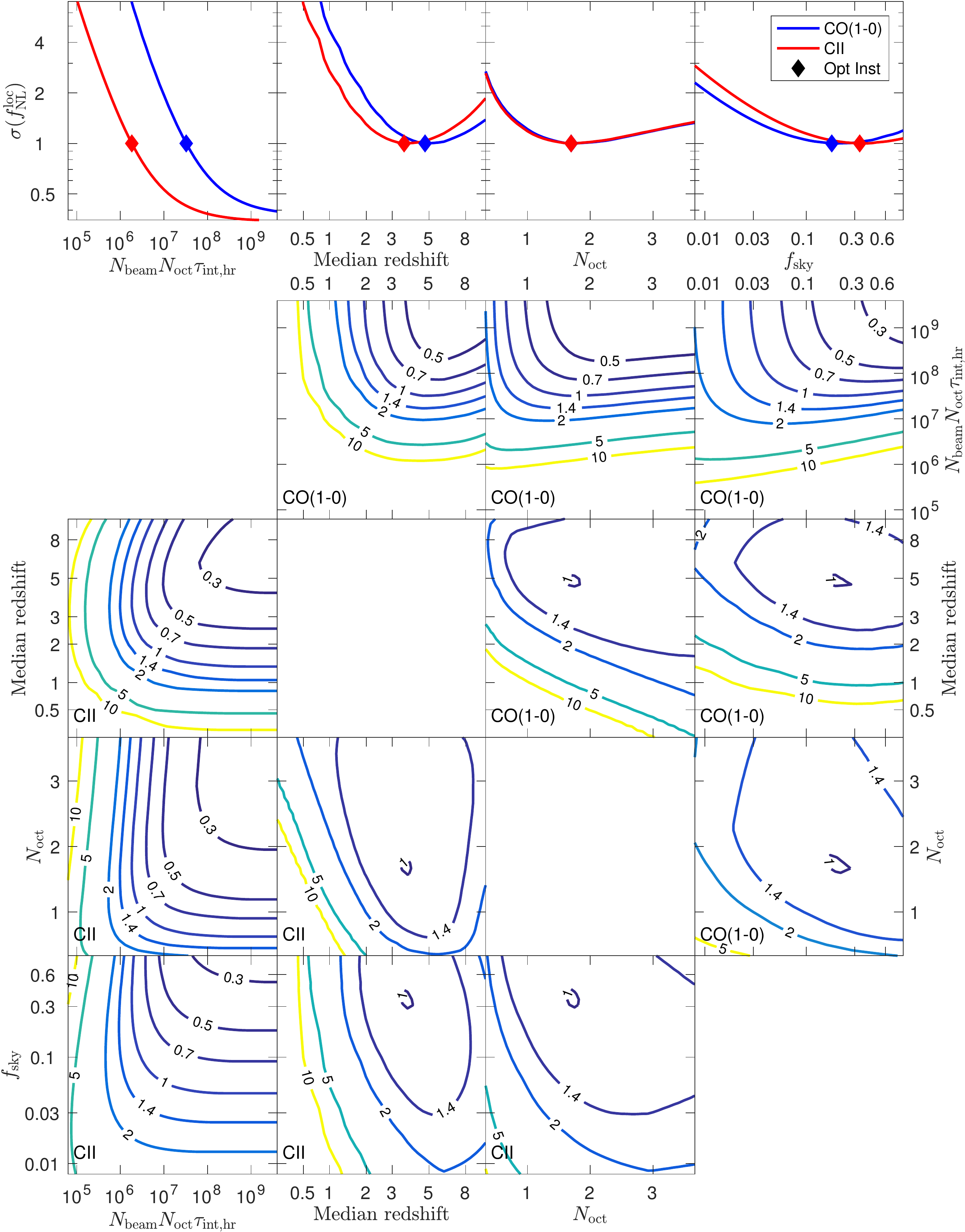}
\caption{The effects of adjusting the four survey parameters on our optimized surveys. \emph{Top row:} The value of $\sigma(f_{\rm NL}^{\rm loc})$ achieved when adjusting a single parameter (and holding all others constant) for the optimized CO (blue) and CII (red) intensity mapping surveys, with respect to the survey parameter values (diamonds) required for the $\sigma(f_{\rm NL}^{\rm loc})=1$ target, as discussed in Section \ref{sec:results}. We find that the optimized surveys are only moderately sensitive to the input parameters, with changes of 10\% (or more, in the case of $N_{\rm oct}$ and $f_{\rm sky}$) producing negligible changes ($<5\%$) in $\sigma(f_{\rm NL}^{\rm loc})$. \emph{Bottom panels:} The impact of varying two of four parameters simultaneously for CO (shown in plots above the diagonal) and CII (plots below the diagonal). 
\label{fig:inst_params}}
\end{center}
\end{figure*}

\vspace{.1in}
\emph{Redshift and spectral coverage:} For even the most modest of constraints on $f_{\rm NL}^{\rm loc}$, both CO and [CII] intensity mapping require a significant spectral coverage. This can be understood in the context of high-redshift nature of these optimized surveys. At high redshift ($z>2$), the change in comoving distance as a function of fractional change scale-factor, $a$, is smaller than it is at low redshift, which makes accessing the largest length-scales difficult without wide redshift coverage. Adding to this is the loss of modes around $k_{\parallel}=0$, which are presumed to be contaminated by continuum emission. The combination of the two factors requires both PNG-focused CO and [CII] intensity mapping surveys to have more than an octave of coverage in order to efficiently produce meaningful constraints on $\sigma(f_{\rm NL}^{\rm loc})$, such that they offer an improvement on the constraints from \textit{Planck}. However, most optimized survey configurations do not require much spectral coverage beyond this range. For both CO and [CII] intensity mapping, the optimal spectral coverage generally resides between 1.5 and 2 octaves (i.e., a factor of 3-4 in spectral coverage), with the median redshift generally set around $z\approx5$. As shown in Figure \ref{fig:inst_params}, we find that for a survey with a target of $\sigma(f_{\rm NL}^{\rm loc})=1$, said survey is only weakly dependent on the precise redshift coverage, particularly when the number of octaves remains fixed. 

We note the appearance of a small, step-like feature that can be seen in both the median redshift and number of octaves for the optimized surveys, which appears to be an artifact of the redshift binning interval chosen for our analysis ($\log_{10}(\Delta[1+z])=0.1$). As the sensitivity to these parameters is sufficiently small (see Figures \ref{fig:optimal_instrument} and \ref{fig:inst_params}), these artifacts are unlikely to meaningfully affect our analysis. We also note that the up-tick in the optimal spectral coverage around $\sigma(f_{\rm NL}^{\rm loc})\lesssim0.2$ is driven in large part by our assumption that broader spectral coverage allows access to a handful of modes with very low values of $k_{\parallel}$, as accessing these modes becomes more important as one approaches the cosmic variance limit.

\vspace{.1in}
\emph{Sky Coverage:} For the optimized cases shown in Figure \ref{fig:optimal_instrument}, $f_{\rm sky}$ scales almost linearly with the required integration time, such that for a target $\sigma(f_{\rm NL}^{\rm loc})\gtrsim 0.6$, the per-mode remains roughly constant ($P_{\rm N}\sim10^{4} \mu\textrm{K}^2\ \textrm{Mpc}^{3}$ for CO, $P_{\rm N}\sim10^{3} \mu\textrm{K}^2\ \textrm{Mpc}^{3}$ for [CII]). Therefore, most of the gain in constraining power for an optimized experiments comes from having more modes to average over, by way of surveying an increasingly large volume.

\vspace{.1in}
\emph{Minimum recoverable scale:} To verify our assumption in Section \ref{ssec:surv_kmax} that the minimum recoverable scale will have little effect on the constraining power of a given survey, we additionally test the dependence of the spatial and spectral resolution (controlled by $D_{\rm ant}$ and $R$) -- both of which affect the maximum value of $k$ accessible by a given experiment. To test this, we evaluate the marginalized constraint on $\sigma(f_{\rm NL}^{\rm loc})$ when $D_{\rm ant}$ is allowed to vary between 0.3 and 30 meters, and $R$ is allowed to vary between 10 and 1000. Generally speaking, we find that the constraining power of a given survey is minimally affected by changes in spatial and spectral resolution, generating changes on the order of only a few percent; larger losses in constraining power are only when $D_{\rm ant}$ or $R$ reach their respective minima. We further find that these losses only appear significant (i.e., $>10\%$) when the spectral or spatial resolution limits correspond to $k_{\rm max} \lesssim 10^{-2}\ h\,\textrm{Mpc}^{-1}$.
\subsection{Dependence on the astrophysical model}\label{ssec:results_astro_model}
With our optimized survey designs for CO and [CII] defined, we now seek to understand how the constraints on local non-Gaussianity depend on astrophysical modeling, i.e. the choice of specific luminosity, halo mass function and the minimum mass of halos hosting a given line. We also test how our forecast is affected by modeling the line bias. In all the tables below the constraints are obtained for the fiducial value of $f_{\rm NL}^{\rm loc} = 1$, marginalizing over all the other parameters as discussed in Section \ref{sec:forecast}, and imposing the Planck priors. 

In Tables \ref{tab:fnl_LF_CO} and \ref{tab:fnl_LF_CII} we show the dependence of the forecasted constraints on the assumption of CO and CII luminosity. In addition to our base luminosity models, for CO, we considered two models with linear relation between CO luminosity and halo mass as explained in Section \ref{ssec:model_var}. The \citet{2016ApJ...830...34K} model for specific luminosity implies an improved constraint on PNG (about $20\%$ better than our base model), whereas the \citeauthor{Pullen:2012su} model would imply a weaker constraint (a factor of 4 decrease in constraining power), versus what our fiducial model would suggest. For [CII], we considered the four models in \citeauthor{Silva:2014ira} shown as $M_1, M_2, M_3, M_4$. The constraints are the best (worst) for $M_1$ ($M_4$) model. The constraints can vary by nearly a factor of 7, depending on the assumed model of CII luminosity.
\vspace{-.1in}
\begin{table}[htbp!]
\caption{Dependence of the 1-$\sigma$ marginalized constraints on local non-Gaussianity on the modeling of CO luminosity.}
\centering
\begin{tabular}{c c  }
\hline \hline
$L_{\rm CO}(M,z)$ & $\sigma(f_{\rm NL}^{\rm loc})$\\ \hline 
\citeauthor{Li:2015gqa} &  1.00 \\
\citeauthor{2016ApJ...830...34K} & 0.807 \\
\citeauthor{Pullen:2012su} &  3.61 \\ 
\hline
\end{tabular}
\label{tab:fnl_LF_CO}
\end{table}
\vspace{-.1in}
\begin{table}[htbp!]
\caption{Dependence of the 1-$\sigma$ marginalized constraints on local non-Gaussianity on the modeling of [CII] luminosity.}
\centering
\begin{tabular}{c c  }
\hline \hline
$L_{[{\rm CII}]}(M,z)$ & $\sigma(f_{\rm NL}^{\rm loc})$\\ \hline 
M1 & 1.00 \\
M2 &  1.22 \\
M3 &  4.08\\ 
M4 &  6.75  \\
\hline
\end{tabular}
\label{tab:fnl_LF_CII}
 \end{table}

In Table \ref{tab:fnl_MF}, we show the dependence of the forecasted constraint on the assumption of the mass function. In addition to Sheth-Tormen (ST) considered in our base model, we consider Press-Schecter (PS) and Tinker (TR) functions. The choice of the mass function affects the constraints from CII and CO similarly. Using the PS function degrades the constraint on PNG by a few percent with respect to the ST function, while using TR function degrades the constraints by nearly a factor of 3.
\begin{table}[htbp!]
\caption{Dependence of the 1-$\sigma$ marginalized constraints on local non-Gaussianity from CO and [CII] on the assumption of halo mass function.}
\centering
\begin{tabular}{c c c  }
\hline \hline
$dn/dM$ 	& $ {\rm CO} : \sigma(f_{\rm NL}^{\rm loc})$ & ${\rm [CII]}: \sigma(f_{\rm NL}^{\rm loc})$  \\  \hline 
PS & 1.09 & 1.02  \\
ST & 1.00 & 1.00 \\
TR10 &  2.74  & 2.65
 \\
\hline
\end{tabular}
\label{tab:fnl_MF} \end{table} 
 
In Table \ref{tab:fnl_M_min} we show the dependence of the forecast of $f_{\rm NL}^{\rm loc}$ on the assumed value of minimum mass of halos contributing to CO and [CII] intensity. In general, the assumption of $M_{\rm min}$ affects the constraints from CO more than that from CII. Compared to our fiducial model with $M_{\rm min} =10^9 M_\odot $, the constraints on $f_{\rm NL}^{\rm loc}$ is degraded by about $26\%$ for CO and $7\%$ for CII when assuming $10^{10} M_\odot$. Assuming $M_{\rm min} = 10^{11}M_\odot$ significantly degrades the constraints (by about a factor of 2 for CO and $40\%$ for CII). While the loss of constraining power at the highest choice of $M_{\rm min}$ is significant for CO, we note that some of this loss can be mitigated by a different choice of redshift range. As shown in Figure \ref{fig:Tbar_line}, the optimal range for the $M_{\rm min}=10^{9}M_{\odot}$ extends to a much higher redshift than it does for $M_{\rm min}=10^{11}M_{\odot}$.

That the constraints are more sensitive to the choice of $M_{\rm min}$ for CO than [CII] can be understood in part as a by-product of our models: the highest mass halos contribute more to total [CII] luminosity versus CO luminosity, as the slope of the CII luminosity for $M_1$ model is shallower than that for CO, particularly for masses of $M_{\rm halo}>10^{12} M_{\odot}$. These higher-biased halos therefore better compensate for the ``loss'' of the lower mass halos in the [CII] model versus the CO model. We note however that this makes the [CII] more sensitive to assumptions about $M_{\rm max}$: setting $M_{\rm max} = 10^{12} M_\odot$, we find that the CII constraints degrade by approximately $20\%$, whereas the constraints on CO only degrade by about $5\%$.

\begin{table}[htbp!]
\caption{Dependence of the 1-$\sigma$ marginalized constraints on local non-Gaussianity from CO and [CII] on the assumption of minimum halo mass $M_{\rm min}$.}
\centering
\begin{tabular}{c c c  }
\hline \hline
$M_{\rm min}$ 	& $ {\rm CO} : \sigma(f_{\rm NL}^{\rm loc})$ & ${\rm [CII]}: \sigma(f_{\rm NL}^{\rm loc})$  \\  \hline 
$10^9 M_\odot$ &  1.00  & 1.00 \\
$10^{10} M_\odot$ & 1.26 & 1.07 \\
$10^{11} M_\odot$  & 2.07 & 1.39
 \\
\hline
\end{tabular}
\label{tab:fnl_M_min}
\end{table}

In Table \ref{tab:fnl_bias}, we show the constraints on local shape for our base astrophysical model when fixing the line bias to that in Equation \eqref{eq:line_bias}, or when considering one free bias parameter for each redshift bin and marginalizing over them. Allowing for biases to be float, marginally degrades the constraint. We note that when assuming the bias to be given by Equation \eqref{eq:line_bias}, we account for the dependence of the bias on cosmological parameters when marginalizing, such that the bias is not completely fixed in this case.
\begin{table}[htbp!]
\caption{Dependence of the constraints on local non-Gaussianity from CO and [CII] on assumption of bias.}
\centering
\begin{tabular}{c c c }
\hline \hline
$b(z)	$ & $ {\rm CO} : \sigma(f_{\rm NL}^{\rm loc})$ & ${\rm [CII]}: \sigma(f_{\rm NL}^{\rm loc})$  \\  \hline 
bias dep on cosmo params  & 1.00 & 1.00\\
5 floating biases  & 1.03  &  1.03\\
\hline
\end{tabular}
\label{tab:fnl_bias}
\vspace{.2in}
\end{table}

Finally, we note that some theoretical models (e.g., \citealt{Li:2015gqa,2017ApJ...846...60C}) remove the enhancement in power from the scatter, by effectively normalizing the power spectrum by the $p_1,\sigma$ term from Equation \ref{eq:scatter}. This is particularly important for the $\sigma_{\rm SFR}$ scatter term found in \citeauthor{Li:2015gqa}, as the halo mass to SFR correlation from \citet{Behroozi:2012sp} reports the mean SFR across halos of a given mass, thus any increase in that mean induced by the logarithmic scatter should be corrected for. However, the same is not necessarily true for the other scaling relationships, particularly those that utilize logarithmic fits to data, like what appears to have been used in the $L_{\rm IR}-L_{\rm CO}$ correlation adopted from \cite{Carilli:2013qm}. As we have parameterized the scatter between halo mass and line luminosity with a single aggregate parameter for CO and [CII] ($\sigma_{\rm CO}$ and $\sigma_{\rm CII}$, respectively), normalizing by $p_1,\sigma$ is likely to underestimate the power spectrum, given the scaling relationships used. In our nominal model, we have therefore chosen not to normalize by this term, but here we evaluate the impact of doing so. In normalizing the power spectrum by $p_{1,\sigma}$, we find that the expected constraints increase to $\sigma(f_{\rm NL}^{\rm loc})=1.45$ for CO-optimized experiment, and $\sigma_{\rm NL}^{\rm loc}=1.54$ for the [CII]-optimized experiment. While significant, we note that this difference is well within the range of the astrophysical models that we have considered in this work (as shown in Tables \ref{tab:fnl_LF_CO} and \ref{tab:fnl_LF_CII}).
\subsection{Constraints on other shapes of PNG}\label{ssec:other_shapes}
For the optimal survey specifications, which achieve $\sigma(f_{\rm NL}^{\rm loc}) = 1$, we also forecast the constraints on the equilateral, orthogonal, and quasi-single shapes of PNG. In Table \ref{tab:fnl_base} we show the constraints on the amplitude of these shapes, $f_{\rm NL}^{\rm shape}$, in addition to the mass parameter of the quasi-single-field model, $\nu$.
\begin{table}[htbp!]
\caption{1-$\sigma$ marginalized constraints on the amplitude of local, equilateral and orthogonal shapes and the amplitude and the mass parameter of quasi-single-filed non-Gaussianity with CO and [CII] intensity mapping.}
\centering
\begin{tabular}{c c c c c c}
\hline \hline
{\rm line} & $\sigma(f_{\rm NL}^{\rm loc})$ &  $\sigma(f_{\rm NL}^{\rm eq})$ & $\sigma(f_{\rm NL}^{\rm orth})$&  $\sigma(f_{\rm NL}^{\rm qsf})$ & $\sigma(\nu)$\\  \hline 
CO &  1.00 & 125  &  44.9 & 62.1 & 14.8 \\
$[{\rm CII}]$  & 1.00 & 89.8 & 39.6 & 45.9 &  11.7 \\
\hline
\end{tabular}
\label{tab:fnl_base}
 \end{table}
 
Our results indicate that, between the two experiments, the optimized [CII] intensity mapping survey achieves tighter constraints on all shapes. Among the shapes considered, the constraints on equilateral PNG is the weakest, since on the large scales, the non-Gaussian correction to the halo bias is nearly constant. For orthogonal shape, the constraints are weaker than local shape and stronger than equilateral, since on large scales the non-Gaussian bias has a $k^{-1}$ scaling. For quasi-single-field model, the scale-dependence correction to the linear bias on large scales has a scaling of $k^{-1/2-\nu}$. Over the allowed range of the mass parameter $\nu$ in this model, $0\leq\nu\leq3/2$, the scale-dependent correction can have a scaling between $k^{-1/2}$ to $k^{-2}$, the latter being the same as that from local shape. Therefore, for our fiducial value of $\nu = 1$, the large-scale non-Gaussian correction to the line bias has a steeper scale-dependence than for the orthogonal shape. One would thus expect the constraint on quasi-single field model to be better than that for the orthogonal shape. We note that while this expectation is correct for unmarginalized errors, parameter degeneracies, in particular between the $f_{\rm NL}^{\rm qsf}$ and $\nu$ render the marginalized error on $f_{\rm NL}^{\rm qsf}$ larger than $f_{\rm NL}^{\rm orth}$. We also note that imposing the Planck priors on cosmological parameters, improves the constraints on local, orthogonal and quasi-single field shapes by a few percent, while for equilateral shape the improvement is about $20\%$. Overall, our results are in broad agreement with forecasted constraints for upcoming galaxy surveys, such as EUCLID and LSST (e.g., \citealt{Sefusatti:2012ye, MoradinezhadDizgah:2018ssw, Karagiannis:2018jdt}).  

The current best constraints on local, equilateral and orthogonal shapes is from measurements of temperature and polarization bispectra of CMB by Planck satellite \citep{2016A&A...594A..17P} which achieved $\sigma(f_{\rm NL}^{\rm loc}) \simeq 6.5$, $\sigma(f_{\rm NL}^{\rm eq}) \simeq 56$ and $\sigma(f_{\rm NL}^{\rm orth}) \simeq 27$ (having converted the CMB constraints to LSS convention by multiplying the Planck reported constraints by a factor of 1.3).  Therefore, while future line intensity mapping surveys have considerable potential to improve upon the current constraints on local shape, improvements on other shapes via power spectra measurements alone will likely prove to be challenging. However, constraints on these other shapes can potentially be improved via measurement of higher-order correlation functions, in particular the bispectrum, which we have not considered here, nor optimized our hypothetical surveys for. Further analysis is required to obtain a more complete estimate of the potential for intensity mapping to constrain these other shapes, which we leave to a future study.  

\section{Discussion}\label{sec:discussion}
\subsection{The Impact of Foregrounds}\label{ssec:disc_max_scale}
\begin{figure*}[!ht]
\begin{center}
\includegraphics[width=.9\textwidth]{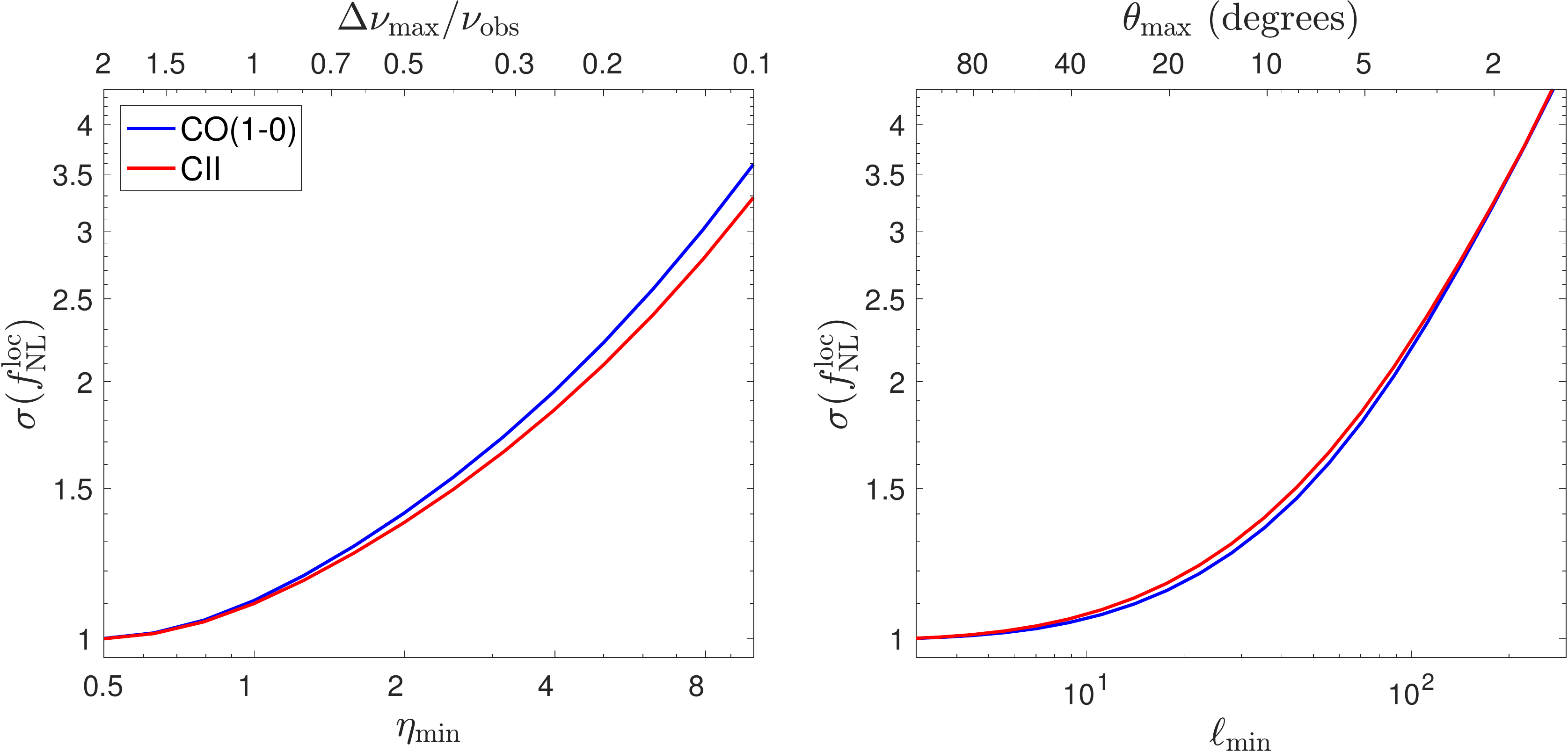}
\caption{\emph{Left:} The impact of screening out the largest-scale $k_{\parallel}$ modes from our optimized surveys, where modes with $\eta<\eta_{\rm min}$ are down-weighted. The achieved value of $\sigma(f_{\rm NL}^{\rm loc})$ in our optimized surveys increases steadily with increasing $\eta_{\rm min}$, further highlighting the importance of recovering the longest-scale modes along the line of sight. \emph{Right}: The impact of  screening out the largest-scale $k_{\perp}$ modes from our optimized surveys, down-weighting modes whose size corresponds to larger angular scales than some given value for $\theta_{\rm max}$. For both CO and [CII], $\sigma(f_{\rm NL}^{\rm loc})$ is only marginally effected until $\theta_{\rm max} \approx 20$ degrees, at which point there is a significant reduction in constraining power due to the loss of most modes with wavenumber $k\lesssim10^{-2.5}\ h\,\textrm{Mpc)}$.
\label{fig:kmin_effect}}
\end{center}
\end{figure*}
It is important to recognize that in modeling the sensitivity of our hypothetical surveys, we have made several optimistic assumptions about the impact of continuum foregrounds: we assumed that foregrounds are smooth over the entirety of survey area (and do not significantly vary in shape as a function of sky position), and that the residuals from these foregrounds are sub-dominant to the spectral line of interest. The modes most likely to show contamination beyond what we have accounted for in Section \ref{ssec:surv_kmax} are the largest-scale modes, where the enhancement in power from PNG is likely to be greatest \citep{Switzer:2017kkz}. To further examine the impact of losing these modes on a potential measurement, we evaluate the impact of increasing the values of $\eta_{\rm min}$ in Equation \eqref{eq:kpar_min} and $\theta_{\rm max}$ in Equation \eqref{eq:kperp_min}. 

Shown in Figure \ref{fig:kmin_effect} is the results of this analysis, with similar effects seen for both the CO and [CII] optimized surveys. In evaluating the impact of $\eta_{\rm min}$, we estimate a loss of $10\%$  in constraining power i.e., $\sigma(f_{\rm NL}^{\rm loc})>1.1$, when $\eta_{\rm min}\approx1.0$, and a loss of $50\%$ when $\eta_{\rm min}\approx2.5$. This would be equivalent to the assumption that the combined foreground and instrumental is smooth over an octave an half an octave, respectively. Similarly for $\theta_{\rm max}$, we find a $10\%$ loss in constraining power when $\theta_{\rm max} \approx 25^{\circ}$, and a $50\%$ loss when $\theta_{\rm max}=8^{\circ}$. 

It is worth noting that for both CO and [CII], the mean brightness of the continuum foregrounds in question can be of order $10^{3}-10^{6}$ times stronger than that of the signal of interest \citep{Switzer:2017kkz,2015ApJ...814..140K}. For CO, the strongest foregrounds are Galactic synchrotron emission, radio point sources, and CMB anisotropies \citep{Planck:2016XI,LaPorta:2008,2015ApJ...814..140K}, while for [CII], the strongest foregrounds are likely to originate from Galactic dust emission and the cosmic infrared background \citep{Planck:2011XVIII,Switzer:2017kkz}. In both cases, for the relatively wideband configuration of our optimized surveys, the amount of power in the $k_{z}=0$ sheet can be as high as $10^{7}-10^{10} \mu\textrm{K}^2\ \textrm{Mpc}^{3}$, which is a factor of $\sim10^{6}$ higher than the shot power for the lines of interest (which scales approximately with total bandwidth per redshift window). This would argue that in order for these contributions to be negligible (i.e., sub-dominant to the shot power), one needs to effectively isolate this contribution to one part in a million in the power spectrum domain. Both \cite{2015ApJ...814..140K} and \cite{Switzer:2017kkz}, have demonstrated, for CO and [CII] respectively, that this level of isolation is possible via rejection of common mode power between frequency channels alone, achieving isolation of one part in $\sim10^{4}-10^{5}$. Both works also suggest that the addition of detailed modeling may be capable of further suppressing contaminant power, although more detailed work would be required for any particular instrument setup.

We note that for both $\ell_{\rm min}$ (the Fourier dual to $\theta_{\rm max}$) and $\eta_{\rm min}$, the point at which there exists a meaningful loss of constraining power is only of order a few times larger than the minimum wavenumber our idealized surveys can reach, as discussed in Section \ref{ssec:surv_kmin}. However, we have made an optimistic assumption for the point-spread function (PSF) in the Fourier domain (effectively, the square of a cylindrical Bessel function), with no consideration for the sidelobes of the PSF causing power from a wider range of $k$ to ``bleed'' into each bin within the 1-D power spectrum. Proper suppression of foregrounds may require more optimized approach than the top-hat style windowing function we have assumed here: e.g., a Gaussian-like windowing function would reduce such bleeding of power, at the cost of increasing $\ell_{\rm min}$ and $\eta_{\rm min}$ by a factor of $\sim\sqrt{2}$, and would translate to a loss of $\approx10\%$ in constraining power for our optimized survey. While beyond the scope of this work, we note that proper accounting of the instrument-specific impact of foregrounds is of critical importance for survey optimization, and defer more detailed work of this issue to future studies.

Beyond continuum foregrounds, we note that we have not explicitly considered the impact of spectral line foregrounds for intensity mapping experiments. While this is not likely to impact intensity mapping experiments targeting the CO(1-0) transition \citep{2016ApJ...830...34K,2017ApJ...846...60C}, the same is not necessarily true for [CII] experiments. In particular, multiple rotational transitions from CO (e.g., CO(3-2), CO(4-3), CO(5-4), CO(6-5)) from moderate redshifts between $z\sim[1-3]$ may dominate over the higher redshift [CII] signal, particularly at $z>6$ \citep{doi:10.1117/12.2057207,Yue:2015sua}. Recent analysis has shown multiple methods to be potentially effective in removing this foreground emission, including masking of known galaxy positions \citep{Sun:2018}, and utilization of the Alcock-Paczynski effect to separate lines emanating from different redshifts \citep{Lidz:2016lub,Cheng:2016}. Rather than attempting to model the losses from a specific foreground suppression method (which is likely to be instrument specific), we consider the cost of foreground \textit{avoidance}, limiting our hypothetical [CII] survey to $z\leq 6$, where the mean [CII] emission is likely to be greater than the aggregate CO emission \citep{Fonseca:2016qqw}. Preserving the amount of spectral coverage would require centering the survey at a redshift of $z=2.9$, and would achieve $\sigma(f_{\rm NL}^{\rm loc}) = 1.05$.
\subsection{Experimental Landscape}\label{ssec:other_experiments}
\begin{table*}[!t]
 \caption{Comparison of hypothetical CO and [CII] intensity mapping surveys}
\begin{center}
\begin{tabular}{l c c c r r r r r}
\hline \hline
Inst Name & line & $z$ & $N_{\rm beam}$ & $N_{\rm oct}$ & $\tau_{\rm int,hrs}$ & $f_{\rm sky}$ & $\sigma(f_{\rm NL}^{\rm loc})$ \\ \hline
COMAP Phase II & CO(1-0) & $2.4-3.4$  & 500 dual-pol   & 0.4 & 10000 & 0.208 & 7.31  \\
ngVLA          & CO(1-0) & $1.4-13.4$ & 214 dual-pol   & 2.6 &  5000 & 0.027 & 2.69  \\ \hline
PIXIE          & [CII]   & $0.2-11.7$ & 1 full-pol     & 3.4 & 10500 & 0.750 & 2.10  \\
OST MRSS       & [CII]   & $0.2-3.2$  & 64 single-pol  & 1.8 &  3000 & 0.100 & 2.21  \\
\hline
\end{tabular}
\label{tab:other_inst}
\end{center}
\end{table*}
Given that the experiments described in Section \ref{sec:results} utilize hypothetical, idealized instruments, we take a moment here to consider the scope of such an instrument, in the context of existing and future, planned instruments. For CO, given the detectors with the sensitivity listed in Section \ref{ssec:instrument_types}, a single octave of spectral coverage, and observing efficiency of 50\%, such a survey would require 2000 single-polarization (or 1000 dual-polarization) receivers a total of $\sim4$ years to complete. For comparison, we note that such an instrument would be only a factor of $\sim2$ times larger (and twice the survey observing length) than the next stage of the CO Mapping Array Pathfinder (COMAP; \cite{Li:2015gqa}.) In fact, if one were able to construct an instrument with instantaneous redshift coverage between $z=[2.2-9.2]$, the required instrument would only require $N_{\rm beam}=552$ dual-polarization beams in order to reach the required survey depth in 4 years time  -- nearly identical to the planned $N_{\rm beam}=500$ planned for Phase II of the COMAP experiment. We note that one critical limitation for COMAP (as currently designed) for performing PNG experiments is the redshift coverage of the instrument -- with less than half an octave of spectral coverage, the final constraint that one might produce with an optimized survey using a Phase II-like instrument is $\sigma(f_{\rm NL}^{\rm loc})=7.31$. In contrast, we consider a large-scale, 5000 hr program on a variant of the Next-Generation Very Large Array\footnote{\url{http://ngvla.nrao.edu/}} (ngVLA), a centimeter-wave interferometer array proposed for deployment in 2034. The aforementioned variant assumes the use of an ultra-wideband receiver with continuous spectral coverage between $8-48$ GHz \citep{2018IAUS..336..426M,Velazco:2017}, with total power measurement capabilities. With only 214 independent dual-polarization beams, and a factor of 2 less in integration time, our hypothetical ngVLA survey would produce a constraint of $\sigma(f_{\rm NL}^{\rm loc})=2.69$; a factor of three improvement on the COMAP Phase II survey, and most significantly limited only by the per-mode noise. The difference in constraining power between these two experiments highlights the need for large bandwidths with intensity mapping surveys to produce meaningful measurements of the local shape of PNG. 

For [CII], given the detector sensitivity found in Section \ref{ssec:instrument_types}, and similarly assuming $50\%$ observing efficiency, such a survey would require a camera with approximately $1.1\times10^{4}$ detectors ($N_{\rm beam}=62$ with 120 spectral channels per beam) 4 years to complete. Such an instrument would have an order of magnitude more detectors than the SPIRE Instrument on the \textit{Hershel} Space Observatory\footnote{\url{http://herschel.cf.ac.uk/mission/spire}}, but two orders of magnitude fewer detectors what is being discussed in the design of the \textit{Origins} Space Telescope\footnote{https://asd.gsfc.nasa.gov/firs/} \citep{Pilbratt:2010hso,Bradford:2018mrss}. For further comparison, we consider first the Primordial Inflation Explorer (PIXIE; \citet{2011JCAP...07..025K}, a spaced-based CMB instrument, utilizing a Fourier-transform Spectrometer (FTS) with spectral coverage between 30 GHz and 6 THz. Due to the low frequency-resolution of PIXIE (fixed at 15 GHz), we limit our consideration to the frequency range to that above 150 GHz. With a proposed 4-year survey, we find that CII intensity mapping with PIXIE would produce a constraint of $\sigma(f_{\rm NL}^{\rm loc})=2.1$\footnote{We note that this value is different than that recorded in \citet{MoradinezhadDizgah:2018zrs} due to significant differences in instrument modeling, and the accounting of for loss of modes as discussed in Section \ref{ssec:surv_kmin}} . We also consider the \textit{Origins} Space Telescope\footnote{\url{https://asd.gsfc.nasa.gov/firs/}} (OST), a spaced-based FIR telescope proposed for deployment in the 2035. More specifically, we evaluate the capability of the Medium Resolution Survey Spectrometer (MRSS) instrument, with wavelength coverage between $30-660\ \rm \mu m$ \citep{Bradford:2018mrss}. With an integration time of  $\tau_{\rm int,hr}=3000$, we find that an optimized survey with \textit{Origins} could achieve a constraint of $\sigma(f_{\rm NL}^{\rm loc})=2.21$:  competitive with other single-tracer constraints with surveys of the next decade. Similar to COMAP, the limiting factor for PIXIE is due limited sensitivity to the CII signal at high redshift, such that the there is little constraining power from $z\gtrsim3$. For \textit{Origins}, the limiting factor is both redshift and sky coverage. An MRSS-like instrument capable of performing a full-sky survey, or a broader redshift range, would be capable of improving upon the above constraints by a factor of two; an instrument capable of both could achieve $\sigma(f_{\rm NL}^{\rm loc})=0.73$.

Table \ref{tab:other_inst} summarizes the survey specifications and the expected constraints on the amplitude of local non-Gaussianity for the four mentioned discussed in this section.

\subsection{Alternative Survey Strategies}\label{ssec:reduced_inst}
Our focus up to this point has been on surveys that are capable of achieving $\sigma(f_{\rm NL}^{\rm loc})=1$, in part as theoretical work suggests that general relativistic effects ought to produce a power spectrum feature similar to what would be observed with the scale-dependent bias associated with $f_{\rm NL}^{\rm loc}\sim 1$, providing a natural ``floor'' for an experiment driven towards a first detection. However, as discussed in Section \ref{ssec:other_experiments}, the scope of such an instrument for both CO and [CII] is generally beyond the scope of currently planned intensity mapping experiments. Moreover, upcoming galaxy surveys such as DESI, LSST, and Euclid are expected to improve the constraints on $f_{\rm NL}^{\rm loc}$ via measurements of galaxy power spectrum and bispectrum. Measurement of scale-dependent bias from galaxy power spectrum is forecasted to achieve $\sigma(f_{\rm NL}^{\rm loc}) = 3.9$ with EUCLID and $\sigma(f_{\rm NL}^{\rm loc}) = 1.4$ for LSST \citep{MoradinezhadDizgah:2017szk}), and $\sigma(f_{\rm NL}^{\rm loc}) = 4.8$ with DESI \citep{Gariazzo:2015qea}. We therefore consider what scope of instrument would be required to achieve $\sigma(f_{\rm NL}^{\rm loc})=3$, of similar order to the single-tracer constraints of the aforementioned galaxy surveys, which may potentially be constructed on a more timely scale than that required achieve they survey outlined in Section \ref{sec:results}.

For CO, such an instrument would require $3.3\times10^{6}$ BOHs, with redshift coverage between $z=[3.0-9.0]$ and $f_{\rm sky}=0.034$ -- approximately 1400 sq. degrees. With a single receiver capable of covering this redshift range, and assuming a 2-year survey (with 50\% observing efficiency), the instrument required to reach this relaxed target would have $N_{\rm beam}=290$. Such an experiment would be well-matched to surveys designed to complement on-going 21-cm epoch of reionization experiments like the Hydrogen Epoch of Reionization Array (HERA; \citet{DeBoer2017}), where cross-correlation with CO(1-0) and CO(2-1) could offer independent verification of the detection of the neutral hydrogen signal \citep{Lidz:2011dx}.

For CII, an instrument of reduced scale would require $2.0\times10^{5}$ BOHs, with redshift coverage between $z=[2.2-7.1]$ and $f_{\rm sky}=0.052$, or roughly 2100 sq. degrees . Given a background-limited set of detectors (as described in \ref{ssec:instrument_types}), a one-year survey (assuming 50\% observing efficiency) with such an instrument would require $N_{\rm beam}=34$, which is of similar scope (half the number of beams, but twice the bandwidth per beam) to the MRSS instrument. We note that our noise-estimates assume an aperture temperature of 40 K, whereas part of the design of \textit{Origins} is the use of an actively-cooled aperture with temperature $\approx6$ K. With an actively cooled aperture, we note that the effective cost of the survey is reduced by nearly an order of magnitude, such that the survey could be completed with an instrument with $N_{\rm pix}\approx7$ (or completed with an MRSS-like instrument in approximately a month of observing).

Several theoretical works (e.g., \citealt{Lidz:2011dx,Gong:2011mf}) have suggested either CO or [CII] as potential tracers to cross-correlate against 21-cm Epoch of Reionization (EoR) experiments. As the optimal redshift window for both CO and [CII] PNG-focused experiments partially overlap with the EoR window ($z\approx[6-10]$), we consider here what scale of instrument would be required to cross-correlate with the HERA \citep{DeBoer2017} which will survey 1440 square degrees, with redshift coverage between $z=[4.7-27]$. As this redshift interval is largely accessible from the ground for [CII], we here consider the requirements for a ground-based instrument, assuming a precipitable water vapor of 0.25 mm (appropriate for winter at Ridge A of the South Pole; \citealt{Lane:1998sp}) for our [CII] optimized experiment. Limiting our consideration to $z=[4.7-10]$, with the same sky coverage, we find that reaching $\sigma(f_{\rm NL}^{\rm loc})=3$ would require $4.1\times10^{6}$ and $6.0\times10^{6}$ BOHs for CO and [CII] respectively. For CO, the instrument requirements are very similar to that required for the optimized $\sigma(f_{\rm NL}^{\rm loc}) = 1$ case. For [CII], the scope of instrument required for undertaking such a survey would be considerable, requiring $N_{\rm beam}=240$ for a 4-year survey with 50\% observing efficiency, with an approximate total of $10^4$ detectors -- approximately an order of magnitude less than that required for the proposed CMB-S4\footnote{\url{https://cmb-s4.org/}} experiment. We note that while this estimate does account for the effective increase in $\eta_{\rm min}$ due to frequency-dependent transmission through the atmosphere, it is at least somewhat optimistic, in that it assumes static atmospheric conditions and instrument performance.

Utilizing multi-tracer techniques \citep{Seljak:2009mlt}, the combined constraining power of both CO and [CII] surveys would be $\sigma(f_{\rm NL}^{\rm loc})=2.1$. While modeling of the HI signal is beyond the scope of this particular work, we note that other theoretical work suggests that the bias for HI ought to be lower than that shown for CO and [CII] here \citep{Li:2017jnt}, suggesting that a combined multi-tracer analysis that utilizes HI, CO, and CII intensity maps could produce constraints on $f_{\rm NL}^{\rm loc}$ of order unity.

While the above estimates were modeled with cross-correlation with the HERA experiment in mind, we note that the above estimates do not account for the high aspect ratio of the survey geometry, which measures more than $20:1$ (with the survey measuring $180^{\circ}\times 8^{\circ}$). Our choice of $\theta_{\rm max}$ (and by extension, $k_{\perp,{\rm min}}$) rested on the assumption of a roughly symmetrical survey area. However, one can account for this by separately defining $\theta_{\rm max,x}$ and $\theta_{\rm max,y}$ (as based on the diameter across the major and minor axes of the survey area) to calculate values for $k_{\rm x,min}$ and $k_{\rm y,min}$. Accounting for this behavior (via modification of Equation \ref{eqn:atten_kmin}), we find that constraining power of a given survey is relatively insensitive to changes in survey geometry \textit{unless} one of the dimensions of the survey drops below $\approx 20^{\circ}$, generally mimicking the behavior shown in Figure \ref{fig:kmin_effect}. Accounting for this additional loss, reaching $\sigma(f_{\rm NL}^{\rm loc})=3$ would require increasing the integration time by at least $\sim30\%$. Under the given scenario, the most efficient approach is to expand the survey area across the shortest dimension of the survey, reducing the number of modes lost at low-$k$ at the cost of limited overlap between the HI and hypothetical CO and [CII] surveys.

\section{Conclusion}\label{sec:conclusion}
Constraints on PNG provide a unique window into the physics of early universe and the origin of primordial fluctuations. In this paper we investigated the potential of future CO and [CII] intensity mapping surveys in constraining PNG. Our focus has been on finding an optimal survey to achieve target sensitivity of $\sigma(f_{\rm NL}^{\rm loc})=1$, and study the impact of variations of astrophysical modeling on the forecasted constraints. We also obtained constraints on other shapes of PNG. Our findings can be summarized as follows:  
\begin{itemize}
\item Achieving $\sigma(f_{\rm NL}^{\rm loc})=1$ with a single tracer (either CO or [CII]) requires an instrument beyond what is planned in current experimental landscape. However, achieving $\sigma(f_{\rm NL}^{\rm loc})=3$, a target in line with expected constraints from power spectrum measurements with upcoming galaxy surveys, may be achievable by an upcoming generation of instruments. With $\sim10^{4}$ hours of integration time, and 1.6 octaves of spectral coverage, a ground-based CO-focused instrument would require a few hundred independent beams, and a space-based [CII]-focused instrument would require a few dozen.
\item For both CO and [CII] intensity mapping experiments, achieving the best possible constraints on $\sigma(f_{\rm NL}^{\rm loc})$ requires broad spectral coverage, optimally covering between a factor of three and five in frequency range. This is due in large part to continuum emission, which will contaminate the longest spatial modes along the line of sight.
\item We studied the impact of various assumptions in the modeling of the line power spectrum on the forecasted constraints on $f_{\rm NL}^{\rm loc}$, including the specific luminosity of galaxies luminous in a given line, halo mass function and minimum mass of halos hosting a given line. Among all, the assumption of specific luminosity of the line has the most significant impact on the constraints on PNG.
\item We also obtained constraints on PNG of several other shapes from a survey that is optimized to probe local PNG. For the survey specifications we chose, achieving improved constraints on orthogonal and equilateral shapes, beyond the limits set by Planck satellite, proves to be challenging. Moreover, constraints on quasi-single-field model are weaker than those forecasted for upcoming galaxy surveys. We note however that the survey specifications that are optimal for constraining the local shape, whose signal is primarily on large scales, are not necessarily optimal for other shapes. Therefore in order to have a more accurate estimation of the potential of line intensity mapping surveys in constraining other shapes, one has to perform a dedicated instrument optimization analysis. Moreover, taking advantage of multi-tracer technique or the bispectrum of line intensity can significantly improve the constraints on all shapes. 
\end{itemize}

\section*{Acknowledgements}
We would like to thank Anastasia Fialkov for collaboration at early stages of this work. We also thank Attila Kovacs for having provided many helpful insights into details related to the survey design portion of this work, as well as his feedback on a draft of this manuscript.  We also thank Eric Switzer, as well as Tzu-Ching Chang, Abigail Crites, and Olivier Dor\'{e}, for fruitful discussions. G.K. would also like to thank the ngVLA Time Domain, Cosmology, and Physics group -- especially Joseph Lazio -- as well as Phillip Bull for their valuable feedback and suggestions in the early stages of this work. A.M.D. is supported under NSF grant AST-1813694.

\bibliography{SCINC} 
\end{document}